\documentclass[a4paper,11pt]{article}
\usepackage{jheppub} 
\usepackage{lineno}
\pdfoutput=1 
\usepackage{subcaption}
\usepackage{graphicx,color}
\usepackage{enumitem}
\usepackage{braket}
\usepackage{slashed}
\usepackage[utf8]{inputenc}
\usepackage{hyperref}
\usepackage{float}
\usepackage{caption}
\usepackage{subcaption}
\usepackage{makecell}
\usepackage{epsfig}
\usepackage{amsmath,amssymb}
\usepackage{hyperref}
\usepackage{diagbox}
\usepackage{lscape}
\usepackage{bbold}
\usepackage{algorithm}
\usepackage{algpseudocode}
\usepackage{amsmath}
\usepackage{amsthm}
\usepackage{comment}
\usepackage{booktabs}
\usepackage{siunitx}
\usepackage{mathrsfs}
\usepackage{soul}

\hypersetup{
	colorlinks=true,
	citecolor=black,
	filecolor=black,
	linkcolor=blue,
	urlcolor=blue
}

\usepackage{dsfont}

\usepackage{dsfont}

\newcommand{\comments}[1]{}

\def\bel#1{\begin{equation} \label{#1}}

\def\ED3{{\scriptscriptstyle ED3}}

\newcommand{\beq}{\begin{equation}}  \newcommand{\eeq}{\end{equation}}
\newcommand{\bal}{\begin{aligned}}   \newcommand{\eal}{\end{aligned}}

\newcommand{\bmat}{\left(\begin{array}}
\newcommand{\emat}{\end{array}\right)}

\newcommand{\cN}{\mathcal{N}}

\newcommand{\cI}{\mathcal{I}}

\newcommand{\cR}{\mathcal{R}}

\newcommand{\cM}{\mathcal M}

\usepackage{comment}
\usepackage[normalem]{ulem}

\newcommand{\I}{\text{i}}

\newcommand{\kom}{\, ,\quad }

\newcommand*{\p}{\mathop{}\!\mathrm \partial}

\usepackage{orcidlink}

\preprint{LITP-26-05}

\title{\boldmath Parameter compression in the flux landscape}

\author[a]{A. Chauhan$\,$\orcidlink{0009-0009-0596-0928},}
\author[b,c]{M. Cicoli$\,$\orcidlink{0000-0003-1709-5651},}
\author[d]{S. Krippendorf$\,$\orcidlink{0000-0001-6374-6828},}
\author[a,e]{A. Maharana$\,$\orcidlink{0000-0001-8627-6398},}
\author[f]{P. Piantadosi$\,$\orcidlink{0009-0000-5284-9485},}
\author[g]{\\A. Schachner$\,$\orcidlink{0000-0002-7287-1476}}

\affiliation[a]{\footnotesize Harish-Chandra Research Institute, 
A CI of Homi Bhabha National Institute, Chhatnag Road, Jhunsi, Allahabad, India 211019.}
\affiliation[b]{\footnotesize Dipartimento di Fisica e Astronomia, Università di Bologna, via Irnerio 46, 40126 Bologna, Italy}
\affiliation[c]{\footnotesize INFN, Sezione di Bologna, viale Berti Pichat 6/2, 40127 Bologna, Italy}
\affiliation[d]{\footnotesize University of Cambridge, Cavendish Laboratory and DAMTP, Cambridge CB3 0WA, United Kingdom}
\affiliation[e]{\footnotesize Leinweber Institute for Theoretical Physics, Randall Laboratory of Physics
University of Michigan, Ann Arbor
450 Church St, Ann Arbor, MI 48109-1040, USA}
\affiliation[f]{New York University Abu Dhabi, PO Box 129188, Saadiyat Island, Abu Dhabi, UAE }
\affiliation[g]{\footnotesize Department of Physics, Cornell University, Ithaca, NY 14853, USA}

\emailAdd{amanchauhan@hri.res.in}
\emailAdd{michele.cicoli@unibo.it}
\emailAdd{slk38@cam.ac.uk}
\emailAdd{anshumanmaharana@hri.res.in}
\emailAdd{pellegrino.piantadosi@nyu.edu}
\emailAdd{as3475@cornell.edu}

\abstract{
We present a data-driven investigation of the exhaustive ensemble of no-scale type IIB flux vacua constructed in \cite{Chauhan:2025rdj}. Using a combination of linear and non-linear dimensionality-reduction techniques, we analyse both flux and moduli spaces and demonstrate that the effective dimensionality of the underlying 12-dimensional flux space is substantially reduced. A central component of our study is a physics-informed autoencoder, which provides a non-linear compression of the flux and moduli data into a low-dimensional latent space. The learned latent representation organises vacua according to desired features and, in particular, isolates distinguished regions associated with small values of the flux superpotential $|W_0|$, revealing non-trivial correlations that are not captured by linear methods. In parallel, we apply tools from topological data analysis, specifically persistent homology, to probe the global structure of the vacuum distribution. This allows us to identify robust, long-lived topological features in both moduli and flux subspaces. This work is a necessary step for developing foundation models in string phenomenology.}

\begin{document}
\maketitle
\flushbottom

\section{Introduction}
\label{sec:intro}

The string landscape \cite{Susskind:2003kw, Douglas:2003um}, consisting of low-energy effective field theories (EFTs) arising from compactifications of string theory, is argued to be finite \cite{Acharya:2006zw,Grimm:2023lrf}, although enormously large \cite{Douglas:2003um}. The vast multiplicity of vacua originates from the multitude of admissible internal compactification manifolds, together with the discrete choices of background fluxes \cite{Bousso:2000xa}. As a consequence, understanding the structure and distribution of vacua across the string landscape has remained a central challenge in string phenomenology.

Traditionally, a systematic analysis of the string landscape was feasible only in highly simplified settings, owing to its immense size and complexity. Early studies, therefore, focused on toy models that capture limited subsets of the full landscape. In recent years, however, advances in software development and machine learning, in particular in generative models, have made more comprehensive analyses increasingly tractable -- enabling previously infeasible data-driven approaches.

Initial applications of optimisation and machine learning methods to decipher the vacuum structure of the landscape include genetic algorithms \cite{Abel:2014xta,Cole:2019enn}, reinforcement learning \cite{Halverson:2019tkf,Krippendorf:2021uxu, Cole:2021nnt,Harvey:2021oue,Abel:2021rrj}, generative models \cite{Erbin:2018csv,Halverson:2020opj,Seong:2024wkt, Krippendorf:2025mhp,Walden:2025cpf}. While these approaches demonstrate the potential of learning techniques to navigate intricate regions of the landscape, they largely operate within restricted sectors or fixed backgrounds.

A principal challenge is to construct methods that enable a global and comparative study of string vacua spanning distinct geometrical settings and their associated low-energy effective descriptions. In practice, solutions of string theory are presented in heterogeneous forms, characterised by moduli spaces of differing dimensionalities and by distinct EFTs, thereby rendering direct comparison non-trivial. Analogous difficulties appear in other scientific domains, for instance, in multi-wavelength astrophysical observations or in data gathered by different particle physics detectors. A compelling strategy to confront such challenges is offered by so-called \emph{foundation models} (see \cite{Hallin:2025ywf} for an overview of existing approaches in fundamental physics), which are engineered to learn across disparate data modalities and to extract common structure from otherwise incommensurate datasets. A central technical component of these approaches is the construction of suitable lower-dimensional representations, typically achieved through compression techniques such as autoencoders (see \cite{Mutter:2018sra} for initial work of compressing heterotic orbifold solutions with autoencoders). In this context, our work may be regarded as a necessary step towards developing such models for theoretical physics and, in particular, for solutions of string theory.

In this work, we demonstrate that autoencoders furnish a powerful and widely applicable framework for compressing the high-dimensional parameter spaces that emerge in flux compactifications. Previous investigations of flux vacua have largely relied on linear dimensionality reduction techniques, as in \cite{Cole:2021nnt}, or on visualisations restricted to carefully selected low-dimensional slices of moduli space, as in \cite{Dubey:2023dvu}. While such methods can provide valuable intuition about selected regions of moduli space, their reliance on linear projections or fixed low-dimensional slices restricts their capacity to faithfully represent the non-linear, and multi-scale structure generated in flux compactifications. By contrast, we show that a physics-informed autoencoder can learn a low-dimensional latent representation of flux space that organises vacua according to target phenomenological characteristics, most prominently the value of the flux superpotential $W_{0}$, in accordance with results in~\cite{Krippendorf:2025mhp}. This latent description exposes non-trivial correlations and geometric structure that remain hidden to linear dimensionality reduction techniques. Importantly, the framework we develop is not tied to a specific Calabi-Yau geometry or a particular flux ensemble; rather, it extends naturally to more general compactifications, higher-dimensional moduli spaces, and datasets that lie beyond the scope of conventional analysis. In this sense, the autoencoder can be used as a conceptual backbone of our programme for exploring the string landscape.

While autoencoders furnish an efficient mechanism for constructing informative low-dimensional representations, a complementary viewpoint is necessary to characterise the global organisation of vacua and to extract structural features that remain stable under changes of embedding or coordinate choice. To this end, we employ methods from topological data analysis (TDA). The use of TDA is particularly well-suited to the study of the string landscape (see~\cite{Cirafici:2015pky,Cole:2017kve} for early work), given the intrinsically high dimensionality and structural complexity of the underlying data. In particular, we apply persistent homology to an exhaustive dataset of Type IIB flux vacua constructed in \cite{Chauhan:2025rdj}. Unlike many machine learning techniques, persistent homology offers a high degree of interpretability: by computing the topology of simplicial complexes built over the dataset, it enables the identification of robust topological features --- such as connected components, loops, and higher-dimensional voids --- that encode information about the organisation of vacua across flux space.

The paper is organised as follows. In Section~\ref{sec:typeiib}, we review the aspects of Type IIB flux compactifications that are pertinent to our analysis. Section~\ref{sec:pca} is devoted to linear dimensionality reduction techniques. In Section~\ref{sec:tda}, we provide a concise introduction to TDA, with the associated results presented in Section~\ref{sec:tda_results_mod} and Section~\ref{sec:tda_results_flux}. Section~\ref{sec:autoencoders} addresses the implementation of autoencoders within this framework. We conclude with a summary and outlook in Section~\ref{sec:summary}.

\section{Type IIB flux compactification}
\label{sec:typeiib}

In this section, we establish our notation and conventions by reviewing the essentials of Type IIB flux compactifications \cite{Giddings:2001yu,Douglas_2007,Grana:2005jc}, with particular emphasis on the no-scale vacua systematically constructed in \cite{Chauhan:2025rdj}.

Let us consider a Calabi-Yau 3-fold $X_3$, which is naturally equipped with a holomorphic 3-form $\Omega$. Upon selecting a symplectic basis $\{\Sigma_{I}, \Sigma^{I}\} \subset H_3(X_3,\mathbb{Z})$, satisfying 
$\Sigma_I \cap \Sigma^J = \delta_I^{\ J}$ and vanishing otherwise, one defines the period vector
\begin{equation}\label{eq:PeriodVecGen}
    \Pi = 
    \begin{pmatrix}
        \int_{\Sigma_I} \Omega \\
        \int_{\Sigma^I} \Omega
    \end{pmatrix}
    =
    \begin{pmatrix}
        \mathcal{F}_I \\
        X^I
    \end{pmatrix}\, ,
\end{equation}
where $X^I$ and $\mathcal{F}_I$ are the periods. The complex structure moduli space is described by special Kähler geometry, such that the $\mathcal{F}_I$ are obtained from a prepotential $F(X)$ via $\mathcal{F}_I = \partial_{X^I}F(X)$. In a patch with $X^0\neq 0$, we introduce affine coordinates $z^i = {X^i}/{X^0}$ with $i=1,\dots,h^{1,2}$. One may use the rescaling freedom of $\Omega$ to fix the gauge $X^0 = 1$. In this gauge, the prepotential reduces to a function $F(z) \equiv F(X)\big|_{X^0=1}$, and the periods take the form
\begin{equation}
    \mathcal{F}_i(z) = \partial_{z^i} F(z)\; , 
    \quad 
    \mathcal{F}_0(z) = 2F(z) - z^i \partial_{z^i} F(z)\, .
\end{equation}

In the large complex structure (LCS) regime, the leading order prepotential is given by
\begin{equation}
    \label{eq:prepotential}
    F(z)=-\frac{1}{6}\widetilde{\kappa}_{ijk}\, z^i\,z^j\,z^k+\frac{1}{2}a_{ij}\,z^i\,z^j+b_i\,z^i +\tilde{\xi}\, ,
\end{equation}
where $\widetilde{\kappa}_{ijk}$ denote the triple intersection numbers of the mirror Calabi-Yau threefold $\widetilde{X}_3$, and $a_{ij}$, $b_i$ are constants fixed by its topological data. In our analysis, we specialise to the example of \cite{Chauhan:2025rdj} with $h^{1,2}=2$, for which the non-vanishing intersection numbers and coefficients are
\begin{equation}
    \widetilde{\kappa}_{111}=9\, , \ \widetilde{\kappa}_{112}=3, \ \widetilde{\kappa}_{122}=1\, ,  \ a=\frac{1}{2} \begin{pmatrix} 9 & 3 \\ 3 & 0 \end{pmatrix}\, , \ b=\frac{1}{4}\begin{pmatrix}
        17 \\ 6
    \end{pmatrix}\, .
\end{equation}

To achieve moduli stabilisation, we introduce quantised background fluxes associated with the three-form field strengths $F_3$ and $H_3$. Expanding these fluxes in the chosen symplectic basis yields two integer-valued flux vectors
\begin{equation}\label{eq:fluxdef}
    f=\left (\begin{array}{c}
                f_{1} \\ 
                f_{2}
                \end{array} 
    \right ) \, ,\quad 
    h=\left (\begin{array}{c}
                h_{1}\\ 
                h_{2} 
            \end{array} 
    \right )  \ \ \ \ \ \ \ \ \ \ \ f_{1},f_{2},h_{1},h_{2}\in\mathbb{Z}^{h^{2,1}+1}\, .
\end{equation}
where the flux quanta are defined as
\begin{equation}
    (f_2)^I 
=\int_{\Sigma_{I}}F_3\, ,\quad (f_1)_I
=\int_{\Sigma^I}F_3\, ,\quad (h_2)^I
=\int_{\Sigma_{I}}H_3\, ,\quad (h_1)_I
=\int_{\Sigma^I}H_3\, .
\end{equation}
The background fluxes induce the scalar potential
\begin{equation}
\label{eq:scalarpotential}
    V_{\text{flux}}= \mathrm{e}^{K} \left ( K^{\tau\bar{\tau}}D_{\tau} W\, D_{\bar{\tau}}\overline{ W}+ K^{i\bar\jmath}D_{i} W\, D_{\bar\jmath}\overline{ W}\right )\, ,\quad D_I W = \partial_I W + (\partial_{I}K) W
\end{equation}
where $K$ and $W$ denote the Kähler potential and flux superpotential of the underlying four-dimensional $\mathcal{N}=1$ supergravity theory,
\begin{equation}
\label{eq:superpotential}
    K= - \ln \left ( -\mathrm{i}\Pi^{\dagger} \cdot \Sigma \cdot \Pi\right )\ - \ln \left ( -\mathrm{i} (\tau - \bar{\tau} )\right )\, ,\quad W= (f-\tau h)^T \cdot \Sigma \cdot \Pi \, .
\end{equation}
We further define the vacuum expectation value of $W$ as
\begin{equation}
\label{eq:superpotvev}
    W_0  = \sqrt{\frac{2}{\pi}}\langle \mathrm{e}^{K}\int G_3\wedge \Omega \rangle\, .
\end{equation}
Supersymmetric vacua arise from flux and moduli configurations satisfying the $F$-term conditions $D_{\tau}W = D_i W = 0$. These are equivalent to the imaginary self-duality (ISD) condition $\star_6 G_3 = \mathrm{i} G_3$ \cite{Giddings:2001yu}. Writing $\tau = c_0 + \mathrm{i} s$, the ISD constraint can be expressed in real form as
\begin{equation}
\label{eq:ISDCond_real} 
    f=\left (s\, \Sigma\cdot \cM+c_{0}\mathds{1}\right )\cdot h
\end{equation}
where the \emph{ISD matrix}
\begin{equation}
\label{eq:ISD_matrix}
    \cM = \left (\begin{array}{cc}
                     \; -\cI^{-1} &\; \cI^{-1}\cR  \\ 
                       \;\;\cR\cI^{-1} &\,  -\cI-\cR \cI^{-1}\cR\;
                       \end{array} \right )\, ,
\end{equation}
is constructed from the real and imaginary parts, $\mathcal{R}$ and $\mathcal{I}$, of the gauge kinetic matrix
\begin{equation}
\label{eq:GaugeKinMatrix}
    \cN_{I J}=\overline{F}_{IJ}+2\I\, \dfrac{\text{Im}(F_{I L})X^{L} \, \text{Im}(F_{J K})X^{K}}{X^{M}\text{Im}(F_{MN})X^{N}}\kom F_{IJ}=\p_{X^{I}}\p_{X^{J}}F\, .
\end{equation}

\section{Principal Component Analysis}
\label{sec:pca}

Having established the framework of Type IIB flux compactifications and the conditions defining supersymmetric vacua, we now turn to a quantitative analysis of the resulting solution space. Even in the relatively simple setting under consideration, the flux space is 12-dimensional and the associated moduli space 6-dimensional, so that the space of vacua is embedded in a high-dimensional ambient space. This makes it difficult to discern global patterns or correlations directly from the raw data.

Our analysis is based on explicit datasets of supersymmetric Type IIB flux vacua constructed in \cite{Chauhan:2025rdj}. These datasets provide exhaustive solutions within the specified region of moduli space and therefore offer a controlled environment in which to investigate the geometric and statistical structure of the landscape.

To systematically probe the structure of the datasets, summarised in Tab.~\ref{tab:summary}, we begin with principal component analysis (PCA). As a linear method, PCA provides a controlled first step: it identifies the dominant directions of variance and offers a transparent characterisation of global correlations among fluxes and moduli. This serves both as an exploratory tool and as a baseline against which more expressive, non-linear methods can be compared. We subsequently move beyond linear structure in Section~\ref{sec:autoencoders}, where autoencoders learn compact latent representations adapted to the intrinsically non-linear structures.

\begin{table}[t!]
    \centering
    \resizebox{1.\textwidth}{!}{
    \begin{tabular}{| c || c | c | c || c | c | c | c |}
         \hline
         & & & & & & &  \\[-1.3em]
         Name & $\text{Im}(z^i)$ & $s$ & $N_{\text{max}}$ & \#$h$ & \#$f$ & \#$(f,h)$ & $\mathcal{N}_{\mathrm{vac}}$ \\[0.2em]
         \hline
         \hline 
         & & & & & & &  \\[-1.2em]
         A & $[2,3]$ & $\bigl [\frac{\sqrt{3}}{2},20\bigl ]$ & $34$ & 82,082 & 1,849,426 & 5,134,862 & 5,140,872  \\[0.3em]
         \hline
         & & & & & & &  \\[-1.2em]
         B & $[2,5]$ & $\bigl [\frac{\sqrt{3}}{2},10\bigl ]$ & $10$ & 1,900 & 6,340 & 12,160 & 12,196  \\[0.3em]
         \hline
    \end{tabular}
    }
\caption{Summary of exhaustive flux vacua datasets for different moduli regions taken from \cite{Chauhan:2025rdj}. 
}\label{tab:summary}
\end{table}

Principal Component Analysis (PCA) is a standard technique for extracting dominant patterns of variation in high-dimensional datasets (see, for example, \cite{Liu:2015mkm,Slatyer:2016qyl}). Given a dataset arranged as a collection of feature vectors, PCA constructs an orthogonal basis of \emph{principal components} by diagonalising the empirical covariance matrix. The eigenvectors define mutually orthogonal directions in feature space, while the corresponding eigenvalues measure the variance captured along each direction. Ordering the components by decreasing eigenvalue yields a hierarchy of directions along which the data exhibit progressively smaller variance. Dimensionality reduction is achieved by projecting the data onto the subspace spanned by the leading components, thereby retaining the maximal amount of variance for a fixed reduced dimension. Further technical details are summarised in App.~\ref{app:pca_comp}.

A well-known limitation of PCA is its sensitivity to the relative scales of the input features: variables with larger numerical spread contribute more strongly to the covariance matrix and may therefore dominate the leading components. In many applications this scale dependence can obscure intrinsic structure. In the present context, however, the situation is more subtle. In flux space, hierarchies between $h$- and $f$-fluxes, as well as among different flux quanta, are intrinsic properties of the ensemble and should not be artificially removed. For moduli vacuum expectation values, by contrast, scale effects can reflect the chosen sampling region rather than structural features of the vacua. For example, in dataset~A the imaginary part of the axio-dilaton satisfies $\mathrm{Im}(\tau)\in[\sqrt{3/2},20]$, and its comparatively large spread causes the leading principal component to align predominantly with the $\mathrm{Im}(\tau)$ direction. This dominance is therefore largely a consequence of the selected domain in moduli space.

We apply PCA to the 12-dimensional integer flux space for the two datasets A and B listed in Tab.~\ref{tab:summary}, both of which arise from exhaustive scans of LCS vacua in \cite{Chauhan:2025rdj}. The variance ratios of the first six principal components are reported in Tab.~\ref{tab:variance_ratios_flux}. These results indicate a substantial concentration of variance in the leading directions, corresponding to an effective reduction of the flux space to approximately five or six dimensions.

\begin{table}[t!]
  \centering
  \begin{tabular}{l *{6}{S[table-format=1.3]}}
    \toprule
    Dataset
     & {PC1} & {PC2} & {PC3} & {PC4} & {PC5} & {PC6} \\
    \midrule
    A & 0.740 & 0.095 & 0.077 & 0.055 & 0.025 & 0.006 \\
    B & 0.638 & 0.187 & 0.087 & 0.043 & 0.031 & 0.011 \\
    \bottomrule
  \end{tabular}
  \caption{The explained variance ratios of the first six principal components for datasets A and B. The concentration of variance in these leading components indicates a pronounced reduction of the effective flux space to approximately five or six dimensions. The associated eigenvectors are provided in App.~\ref{app:pca_comp}.}
  \label{tab:variance_ratios_flux}
\end{table}

\begin{figure}[h!]
    \centering
    \includegraphics[width=1.0\textwidth]{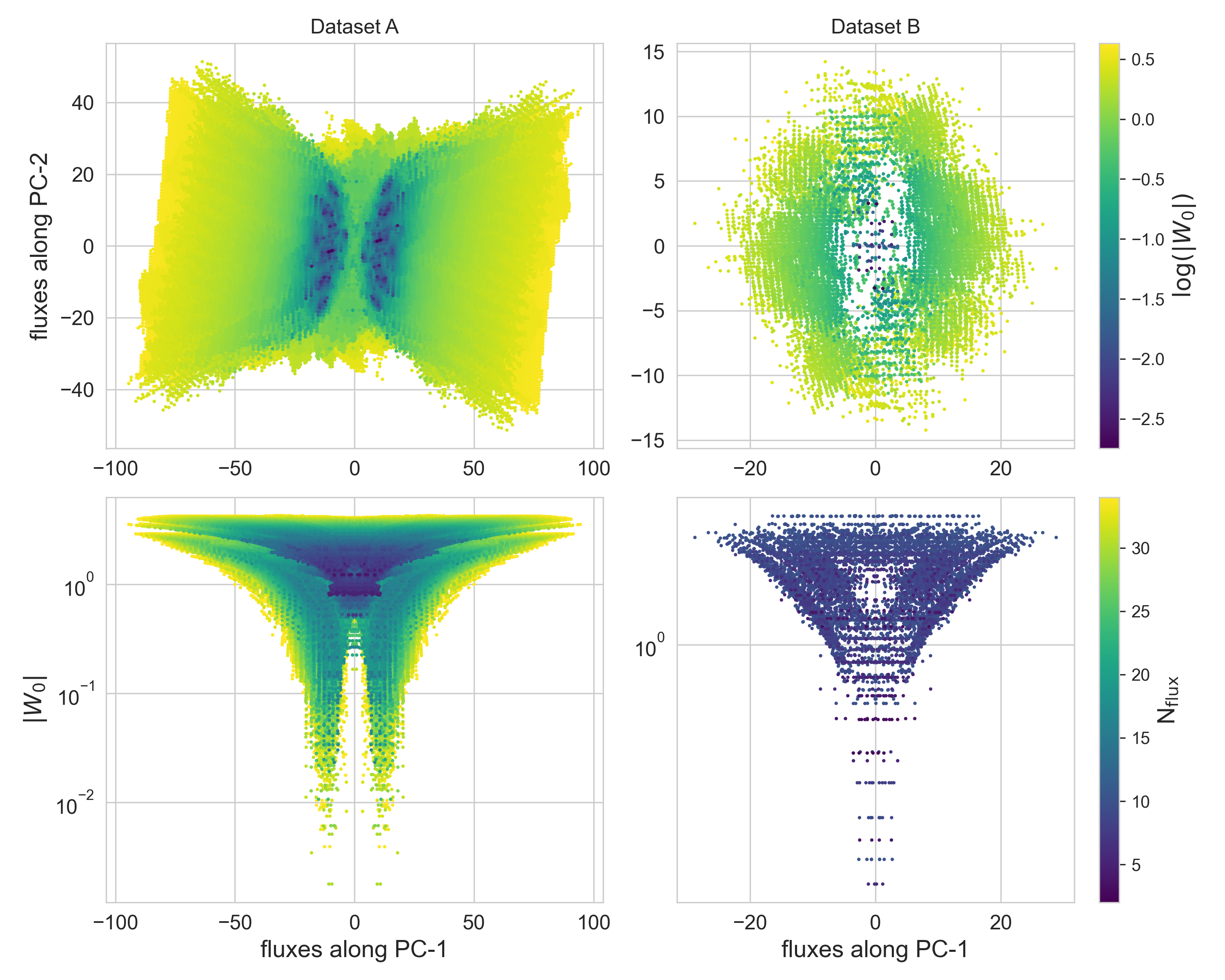}
\caption{Flux configurations projected onto the first and second principal components are shown for dataset~A (left column) and dataset~B (right column). The lower panels, together with the colour coding in the upper panels, display the corresponding distributions of the superpotential along these principal directions.}
    \label{fig:pca_fluxes}
\end{figure}

Fig.~\ref{fig:pca_fluxes} displays the flux vacua projected onto the first two principal components for dataset~A (left) and dataset~B (right). The colour scale in the upper panels represents $|W_0|$, while the lower panels indicate the flux-induced D3-brane charge $N_{\text{flux}}$. Vacua with small $|W_0|$ are concentrated near the origin along the first principal component, whose dominant contribution arises from $(f_1)^2$. In both datasets, the smallest values of the superpotential are restricted to a narrow region around zero. In dataset~A, a mild bifurcation is visible, reflecting additional structure in the larger flux ensemble. These observations indicate that the first principal component captures a substantial fraction of the variance relevant for $|W_0|$, and may therefore serve as a useful direction for targeted searches of vacua with parametrically small superpotential. Moreover, comparison of the lower panels shows that the distribution in dataset~B (bottom right) is embedded within the broader pattern observed for dataset~A (bottom left), pointing towards a degree of universality across ensembles with different values of $N_{\text{max}}$.

\begin{figure}
    \centering
    
    \begin{subfigure}{0.48\linewidth}
        \centering
        \includegraphics[width=\linewidth]{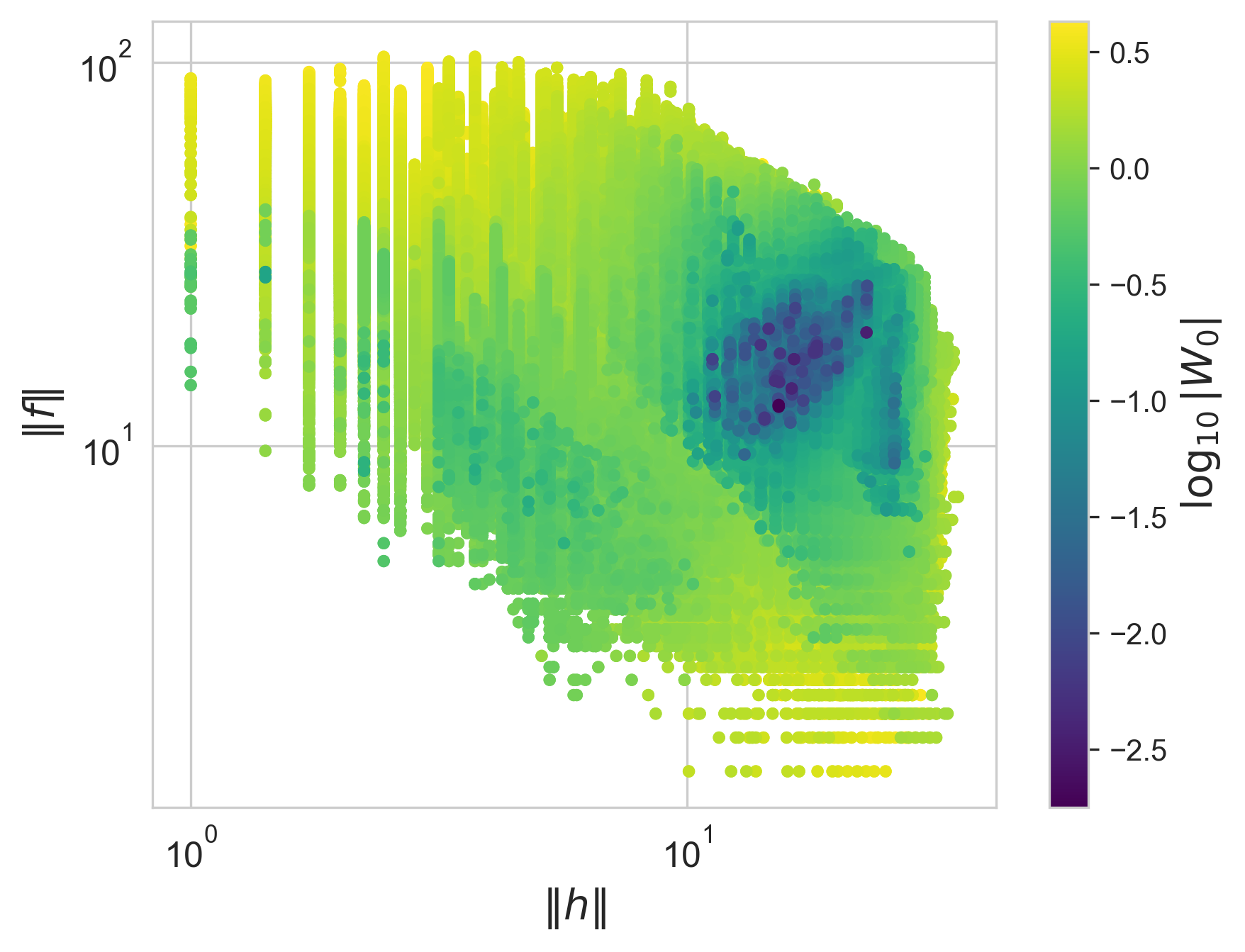}
        \caption{Dataset A}
        \label{fig:flux_norm_A}
    \end{subfigure}
    \hfill
    \begin{subfigure}{0.48\linewidth}
        \centering
        \includegraphics[width=\linewidth]{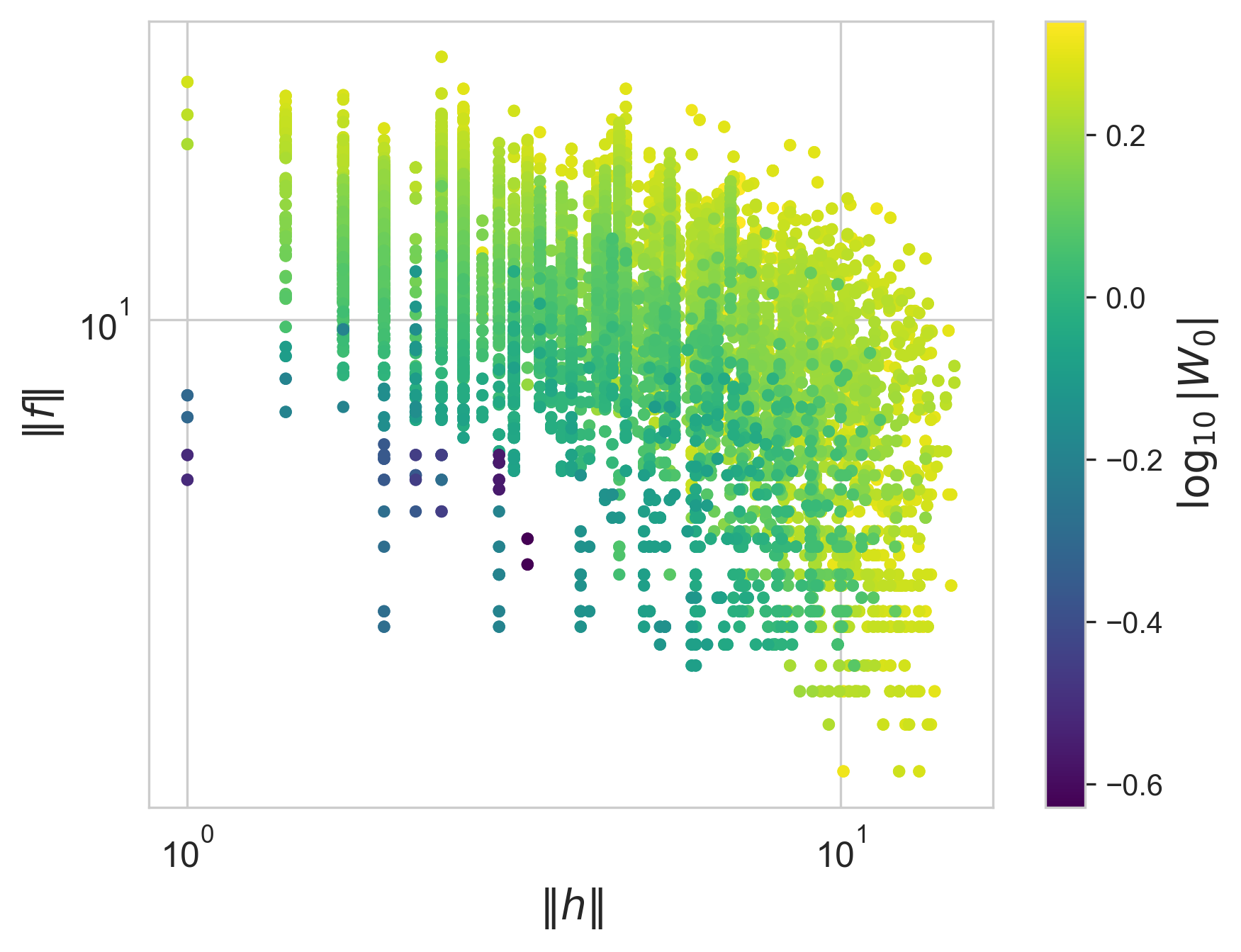}
        \caption{Dataset B}
        \label{fig:flux_norm_B}
    \end{subfigure}
    
    \caption{Scatter plots of the Euclidean norms of the NS--NS and R--R flux vectors, $\|h\|$ and $\|f\|$, for flux vacua in datasets A and B. The points are colour-coded by $\log_{10}|W_0|$.}
    \label{fig:flux_norm_hierarchies}
\end{figure}

Fig.~\ref{fig:flux_norm_hierarchies} shows the distribution of flux vacua in the plane defined by the Euclidean norms of the NS--NS and R--R flux vectors for both the datasets, $\|h\|$ and $\|f\|$, with each point colour-coded by $\log_{10}|W_0|$. In the case of dataset~A, a pronounced structural difference emerges between vacua with superpotential of order unity and those with parametrically small $|W_0|$. Vacua with $|W_0|\sim \mathcal{O}(1)$ occupy a wide region of flux space and typically display marked hierarchies\footnote{The hierarchy between $f$- and $h$-fluxes can be traced back to the structure of \eqref{eq:ISDCond_real}, where the properties of the ISD matrix generically induce asymmetric relations between the two flux sectors.} between the magnitudes of the $f$- and $h$-fluxes. By contrast, vacua with smaller $|W_0|$ are concentrated in a more restricted domain of the $(\|h\|,\|f\|)$ plane, characterised by a substantially reduced hierarchy between the two flux norms. This pattern suggests that achieving small values of the superpotential is correlated with comparatively balanced NS--NS and R--R flux configurations, rather than with large disparities in their overall magnitudes. Exploiting this correlation provides a practical criterion for refining search algorithms aimed at identifying vacua with parametrically small $|W_0|$, namely by preferentially sampling flux configurations with limited norm hierarchies. 

We may refine the analysis by decomposing the flux configurations into the subspaces $f_1$, $f_2$, $h_1$, and $h_2$, which enter differently in the ISD constraint \eqref{eq:ISDCond_real} used to generate the vacua and are therefore expected to exhibit distinct statistical behaviour. Performing PCA separately on these sectors for dataset~A, and inspecting the eigenstructure of the corresponding covariance matrices, reveals that the original 12-dimensional flux space can again be effectively reduced to a five-dimensional subspace. In this basis, the dominant directions are well approximated by the coordinates $(f_1)^1$, $(f_1)^2$, $(f_2)^3$, $(h_1)^1$, and $(h_2)^3$.

In parallel with the flux analysis, we apply PCA to the VEVS of the moduli for both the unstandardised and standardised versions of datasets~A and~B, where by \textit{standardised} we mean that each feature has been mean-centered and scaled to unit variance. The corresponding explained variance ratios are summarised in Tab.~\ref{tab:variance_ratios_roots}. For dataset~A, the first principal component accounts for nearly $98\%$ of the total variance, indicating an effective reduction of the six-dimensional moduli space to an approximately one-dimensional subspace. This dominant direction is largely aligned with $\mathrm{Im}(\tau)$, so that the primary source of variation is controlled by the string coupling $g_s = 1/\mathrm{Im}(\tau)$. Dataset~B, by contrast, displays a richer structure, retaining roughly three effective dimensions as a consequence of its more restricted range of $\mathrm{Im}(\tau)$. We stress that these effective directions correspond to linear combinations of the original moduli rather than to individual fields. For the standardised datasets, all principal components contribute at a comparable level, signalling the absence of a single dominant direction in moduli space.

\begin{table}[ht]
  \centering
  \begin{tabular}{l c c c c c c }
    \toprule
    Dataset & PC1 & PC2 & PC3 & PC4 & PC5 & PC6 \\
    \midrule
A     & 0.980 & 0.004 & 0.004 & 0.004 & 0.004 & 0.004 \\
B     & 0.770 & 0.120 & 0.064 & 0.017 & 0.015 & 0.015 \\
\hline
Std.\ A & 0.176 & 0.168 & 0.167 & 0.166 & 0.165 & 0.159 \\
Std.\ B & 0.239 & 0.179 & 0.161 & 0.161 & 0.157 & 0.104 \\
    
    \bottomrule
  \end{tabular}
    \caption{Explained variance ratios for the moduli space of  datasets~A and~B, together with their standardised counterparts.}
      \label{tab:variance_ratios_roots}
\end{table}

As emphasised above, PCA is intrinsically linear and is therefore limited to identifying global linear correlations in the data. While it provides a useful first characterisation of dominant variance directions, it is not designed to capture more intricate geometric patterns or higher-order correlations that may be encoded in the distribution of vacua within flux or moduli space. To address this limitation, we subsequently employ autoencoders, introduced in Section~\ref{sec:autoencoders}, as non-linear dimensionality reduction tools based on neural network architectures. These models are capable of learning compact latent representations adapted to non-linear manifolds embedded in the high-dimensional flux space. 

Before turning to this non-linear compression framework, however, we complement the PCA analysis with a study of the global topological structure of the datasets. In the next section, we therefore introduce methods from TDA, which enable us to probe robust structural features of the moduli and flux spaces beyond linear variance-based diagnostics.

\section{Topological Data Analysis}
\label{sec:tda}

Topological and geometric concepts are naturally defined for continuous spaces, whereas a finite dataset does not, by itself, carry intrinsic topological information. To associate topology to a dataset $M$, one typically endows it with the structure of a metric space $(M,\rho)$ and constructs a family of simplicial complexes by connecting points whose pairwise distances lie below a given scale. One then studies the evolution of topological invariants of these complexes as the scale parameter varies. 

Topological data analysis (TDA) builds on this construction to provide a largely coordinate-independent characterisation of the data. Our focus is on persistent homology \cite{perea2018briefhistorypersistence}, a framework rooted in algebraic topology that quantifies the multi-scale topological features of a dataset. By tracking features that persist across a range of scales, persistent homology identifies robust geometric and topological structures that are not captured by standard clustering or variance-based dimensionality reduction techniques. Persistent homology has found applications across a wide range of disciplines, including cosmology, statistical analyses and parameter inference \cite{Cole:2017kve, Biagetti:2020skr, Cole:2020gkd,Biagetti:2022qjl,Yip:2024hlz,Calles:2025dli, DES:2025akz}.

\subsection{Persistent homology}
\label{sec:persistence_homology}

In this section, we briefly review\footnote{Readers familiar with these concepts may proceed directly to Section~\ref{sec:tda_results_mod} and Section~\ref{sec:tda_results_flux}.} the basic notions of persistent homology relevant for our purposes.
Given a finite metric space $(M,\rho)$, one may construct a graph by connecting pairs of points whose mutual distance is smaller than a prescribed threshold $\epsilon$, thereby obtaining the associated \emph{$\epsilon$-neighbourhood graph}. While such graphs encode pairwise proximity information, they do not capture higher-order relations among collections of nearby points. TDA refines this construction by introducing higher-dimensional building blocks in the form of simplices.
 
A \emph{$k$-simplex} is determined by $k+1$ affinely independent points 
$v_0, \ldots, v_k \in \mathbb{R}^n$\footnote{Explicitly, a $k$-simplex is the convex hull 
$\sigma = \left\{ \sum_{i=0}^k \lambda_i v_i \;\middle|\; \lambda_i \geq 0, \ \sum_{i=0}^k \lambda_i = 1 \right\}$.}. 
In low dimensions, these correspond to points, line segments, filled triangles, and filled tetrahedra. Higher-dimensional simplices are defined analogously. A \emph{simplicial complex} $K$ is a finite collection of simplices satisfying:  
\begin{itemize}
    \item \textbf{Closure under faces:} If $\sigma \in K$, then every face of $\sigma$ is also in $K$.\footnote{A face of a simplex $\sigma$ is any simplex spanned by a subset of its vertices.}
    \item \textbf{Intersection condition:} If $\sigma, \tau \in K$, then $\sigma \cap \tau$ is either empty or a face of both $\sigma$ and $\tau$. 
\end{itemize}

\noindent Simplicial complexes thus provide a combinatorial model of a topological space, built from elementary simplices and encoding geometric relations in purely combinatorial data. This discrete formulation makes them particularly suitable for algorithmic treatment in TDA. For a finite dataset, several standard constructions of simplicial complexes are available; we briefly recall those relevant for the present work.

\medskip  

\noindent\textbf{Vietoris--Rips Complex.} Let $X $ be a finite set of points in a metric space $(M, d) $, and let $\epsilon > 0 $ denote a filtration parameter. The \emph{Vietoris--Rips complex} $\mathrm{VR}_\epsilon(X) $ is the abstract simplicial complex whose $k$-simplices are subsets $\sigma \subseteq X $ of size $k+1 $ such that 
\[
d(x_i, x_j) \leq \epsilon \quad \text{for all } x_i, x_j \in \sigma.
\]
In other words, a simplex is included if all of its vertices are pairwise within distance $\epsilon$. 

\medskip  

\noindent\textbf{\v{C}ech Complex.}  
The \emph{\v{C}ech complex} $\check{C}_\epsilon(X) $ is defined as the abstract simplicial complex whose $k$-simplices correspond to subsets $\sigma \subseteq X $ of size $k+1 $ such that the closed $\epsilon$-balls centred at the points of $\sigma $ have a non-empty intersection:
\[
\bigcap_{x \in \sigma} B_\epsilon(x) \neq \varnothing,
\]
where $B_\epsilon(x) := \{ y \in M \mid d(x,y) \leq \epsilon \} $.  
By the Nerve theorem, the \v{C}ech complex is homotopy equivalent to the union of $\epsilon$-balls around $X$, and hence captures the correct topology of this union. 

While the \v{C}ech complex has a precise topological interpretation, it is often computationally expensive to construct in high dimensions. The Vietoris--Rips complex, by contrast, is easier to compute since it only depends on pairwise distances. In practice, persistent homology is frequently computed using Vietoris--Rips complexes, with the understanding that they provide an approximation to the \v{C}ech complex as $\mathrm{VR}_{\epsilon}(X) \subseteq \check{C}_\epsilon(X) \subseteq \mathrm{VR}_{2\epsilon}(X)$.

Persistent homology is a central tool in TDA for extracting multi-scale topological information from finite datasets. A key theoretical property is its stability: small perturbations of the underlying point cloud, measured for instance in bottleneck distance, induce correspondingly small changes in the associated persistence diagram \cite{10.1145/1064092.1064133}. This ensures that the extracted topological features are robust under moderate noise or sampling fluctuations. Homology assigns algebraic invariants to a topological space that detect $k$-dimensional holes. In contrast to classical homology, which associates invariants to a fixed space, persistent homology studies how these features evolve along a \emph{filtration}. A filtration is a nested sequence of simplicial complexes $K_0 \subseteq K_1 \subseteq \cdots \subseteq K_n$
typically generated by varying a scale parameter $\epsilon_i$. As $\epsilon_i$ increases, additional simplices are included, and the complex transitions from a sparse configuration of disconnected vertices to an increasingly connected and higher-dimensional structure.

This construction reflects the idea that relevant topological features may only become visible at specific scales. For each $K_i$, one computes the homology groups $H_k(K_i)$, and persistent homology tracks the appearance and disappearance of $k$-dimensional features as $i$ increases. Each feature is assigned a \emph{birth} scale $b$ and a \emph{death} scale $d$, recorded as an interval $[b,d)$. These intervals are commonly visualised in a persistence diagram\footnote{An equivalent representation is given by barcodes, where horizontal bars (grouped by homology degree $H_k$) indicate the birth and death scales of individual features.}, in which each feature corresponds to a point in the $(b,d)$-plane. Points located further from the diagonal $b=d$ represent features that persist over a wide range of scales and are therefore regarded as more significant.

The coordinate-independent and inherently multi-scale character of persistent homology makes it particularly well-suited for analysing complex high-dimensional datasets, such as those arising in the string landscape, where conventional clustering or variance-based dimensionality reduction methods may not adequately capture global structural organisation.

\subsection{Persistent homology on moduli space }
\label{sec:tda_results_mod}

In this subsection, we apply persistent homology both to selected projections and to the full six-dimensional moduli VEVs, with particular emphasis on the computation of $H_1$ homology classes and their persistence across filtration scales. As a first step, it is instructive to analyse lower-dimensional projections onto the complex planes parametrised by $\tau$, $z^1$, and $z^2$, where visual inspection can be directly correlated with topological signatures. Moreover, at tree level the Kähler potential factorises between the complex structure moduli and the axio-dilaton. Since this structure underlies the construction of the flux vacua considered here, it is natural to investigate the $\tau$-plane and the complex structure moduli sectors separately before turning to the full moduli space.

A crucial aspect of our study is that the persistent homology analysis is performed on an exhaustive set of flux vacua, rather than on randomly generated samples or truncated subsets. As a result, the identified topological features correspond to genuine structures of the underlying solution space,
rather than artefacts of incomplete sampling. This point is particularly significant for persistent homology, where the addition or removal of even a single outlier can substantially modify the birth and death scales of long-lived cycles, thereby altering the global appearance of the persistence diagram.\footnote{We thank Gary Shiu for emphasising this point.} By analysing the complete dataset, we avoid spurious distortions in the persistence structure that could otherwise arise from the progressive inclusion of additional vacua. This is in contrast to what has been done in the literature so far for persistence homology applied to the string landscape, as in \cite{Cirafici:2015pky, Cole:2018emh}.

As summarised in Tab.~\ref{tab:summary}, datasets~A and~B comprise $5{,}140{,}872$ and $12{,}196$ vacua, respectively. The computation of persistent homology using the Vietoris-Rips complex is constrained by rapidly increasing memory requirements, since the number of simplices grows combinatorially with the number of points. In practice, this renders calculations beyond $\mathcal{O}(10^4)$ data points and particularly for higher homology groups ($H_2$ and above), computationally prohibitive. For dataset~B, the complete set of points can be employed for the computation of $H_0$ and $H_1$. However, the evaluation of $H_2$ necessitates subsampling, typically at the level of $20\%$ or more of the data, implemented via greedy permutation\footnote{See App.~\ref{app:tda_datA} for more details on this subsampling.} (farthest point sampling) \cite{DBLP:journals/corr/CavannaJS15} as provided in \texttt{Ripser} \cite{ctralie2018ripser}. 

\begin{figure}
    \centering
    \includegraphics[width=1.0\linewidth]{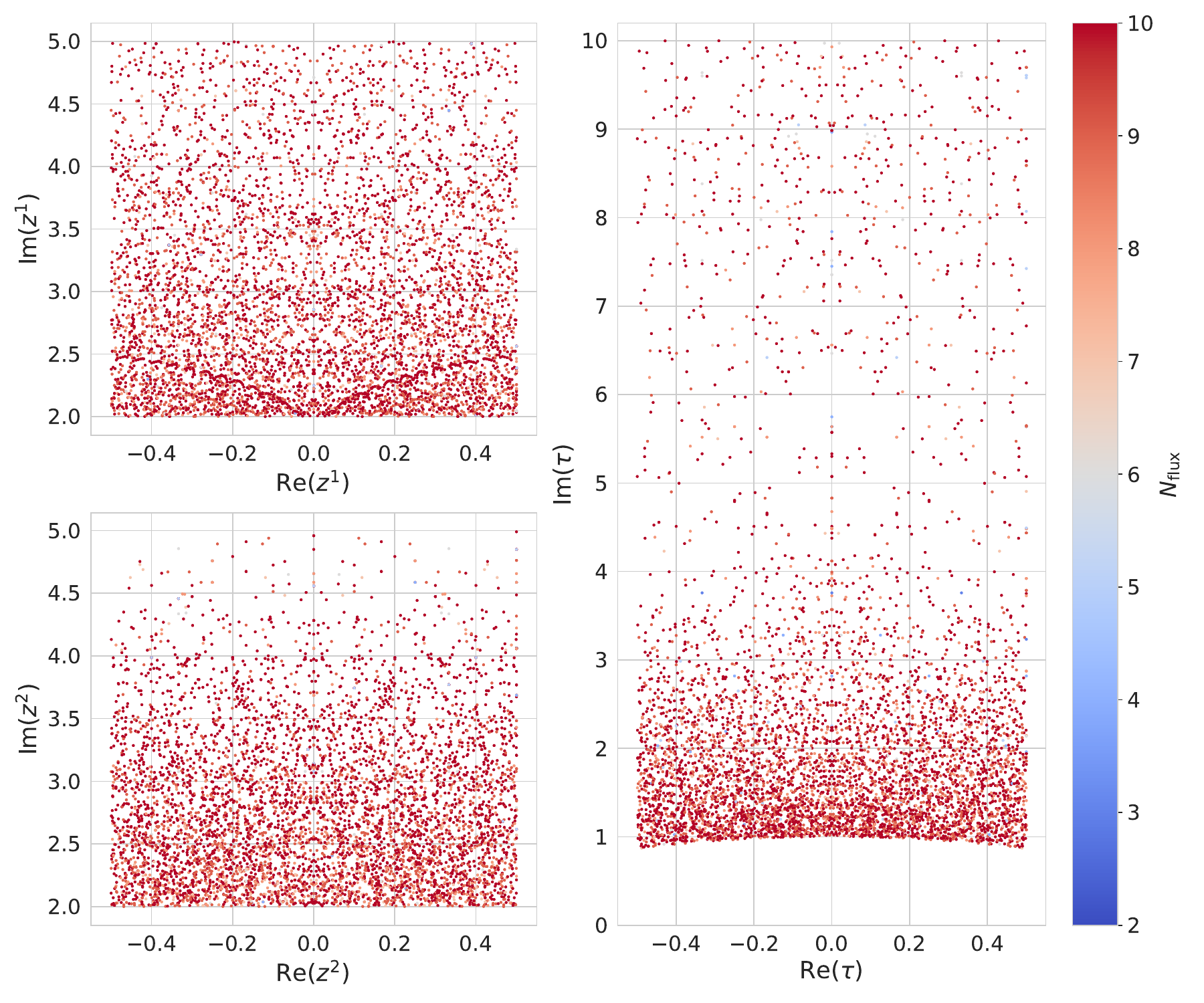}
   \caption{Distributions of the moduli for varying values of the flux-induced D3-brane charge $N_{\text{flux}}$ in dataset~B: (top left) projection onto the $z^1$-plane, (bottom left) projection onto the $z^2$-plane, and (right) projection onto the $\tau$-plane. In each case, the distributions exhibit reflection symmetry about the imaginary axis and display non-trivial geometric patterns, including arc-like structures and closed loops.}
    \label{fig:roots_B}
\end{figure}

Fig.~\ref{fig:roots_B} illustrates the distributions of the moduli $(z^1, z^2, \tau)$ for dataset~B, stratified by the flux-induced D3-brane charge $N_{\text{flux}}$. The resulting vacuum configurations exhibit a pronounced geometric organisation, including arc-like formations and closed loops, signalling non-trivial correlations among the stabilised moduli. A clear trend is visible in the $\tau$-plane: vacua with larger values of $N_{\text{flux}}$ preferentially accumulate at smaller $\mathrm{Im}(\tau)$, corresponding to stronger string coupling. This behaviour reflects a systematic relation between flux quanta and the stabilised value of the axio-dilaton. More generally, the structured nature of these distributions demonstrates that the ensemble of flux vacua is highly organised and already encodes significant geometric information prior to the application of persistent homology.

Before turning to the results, we highlight a discrete symmetry exhibited by our datasets. As is evident in the $\tau$-, $z^1$-, and $z^2$-plane projections shown in Fig.~\ref{fig:roots_B}, the distributions are invariant under the reflections 
$\mathrm{Re}(\tau)\to -\mathrm{Re}(\tau)$ and $\mathrm{Re}(z^i)\to -\mathrm{Re}(z^i)$. 
This symmetry was discussed in \cite{Chauhan:2025rdj} and originates from the behaviour of the $F$-flatness conditions under complex conjugation, as pointed out in \cite{DeWolfe:2004ns}. More precisely, if $(\tau_*,z^i_*)$ is a supersymmetric solution for flux vectors $f$ and $h$, then there exists a second solution at 
$\tau'=-\overline{\tau}_*$ and $z^{i\,\prime}=-\overline{z}^i_*$, 
provided the fluxes transform as 
$f'^T=f^T\cdot U$ and $h'^T=-h^T\cdot U$. 
In our specific example, the matrix $U$ is given by
\begin{equation}
U=\left(
\begin{array}{cccccc}
 1 & 0 & 0 & 0 & 0 & 0 \\
 0 & -1 & 0 & 0 & 0 & 0 \\
 0 & 0 & -1 & 0 & 0 & 0 \\
 0  & 0 & 0 & -1 & 0 & 0 \\
 0 & 9 & 3 & 0 & 1 & 0 \\
 0 & 3 & 0 & 0 & 0 & 1 \\
\end{array}
\right)   \quad \Longrightarrow \quad U\cdot \Sigma\cdot \Pi(-\overline{z}^i_*) = \Sigma\cdot \overline{\Pi(z^i_*)} \, .
\end{equation}
The corresponding $F$-term conditions satisfy
\begin{equation}
D_{\tau}  W (-\overline{\tau}_*,-\overline{z}^i_*) =-\overline{D_{\tau}  W (\tau_*,z^i_*)}\; ,\quad
D_i  W (-\overline{\tau}_*,-\overline{z}^i_*)= -\overline{D_i  W (\tau_*,z^i_*)}\,. 
\end{equation}
The fixed locus of this reflection symmetry is characterised by $\mathrm{Re}(\tau)=\mathrm{Re}(z^i)=0$, which defines a codimension-three submanifold of the full six-dimensional moduli space. Along this locus, the effective topology is correspondingly reduced. Importantly, this symmetry relates physically distinct vacua. This contrasts with the $SL(2,\mathbb{Z})$ and $Sp(6,\mathbb{Z})$ dualities, which act as gauge redundancies mapping physically equivalent configurations to one another.

\begin{figure}
    \centering
    \includegraphics[width=1.0\linewidth]{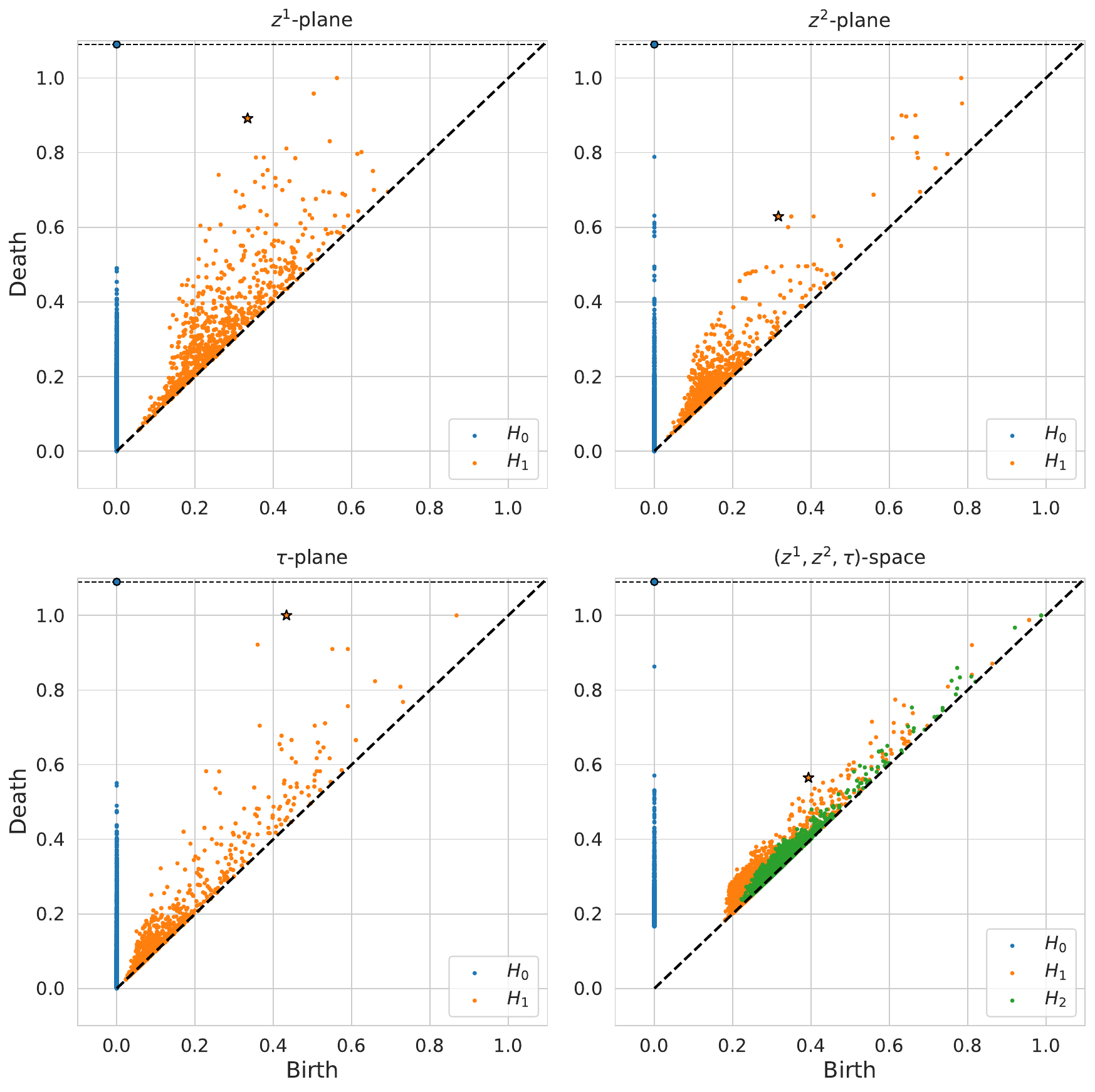}
    \caption{Persistent homology analysis of the vacua in dataset~B: (top left) projection onto the $z^1$-plane, (top right) projection onto the $z^2$-plane, (bottom left) projection onto the $\tau$-plane, and (bottom right) the full six-dimensional $(z^1,z^2,\tau)$ moduli space. Homology classes are colour-coded by degree, with $H_0$ (connected components) shown in blue, $H_1$ (one-cycles) in orange, and $H_2$ (two-cycles) in green. The most persistent $H_1$ class is highlighted by $\star$.}
    \label{fig:PH_roots_B}
\end{figure}

In Fig.~\ref{fig:PH_roots_B}, we display the results of the persistent homology analysis based on the Vietoris--Rips complex constructed from the moduli of dataset~B. The birth and death parameters have been rescaled for each moduli subspace to the interval $[0,1]$\footnote{For each moduli subspace, all finite birth--death pairs $(b_i,d_i)$ are mapped to
\begin{equation*}
    (b_i,d_i)\mapsto 
\left(\frac{b_i-b_{\min}}{d_{\max}-b_{\min}},\,
      \frac{d_i-b_{\min}}{d_{\max}-b_{\min}}\right),
\end{equation*}
where $b_{\min}$ and $d_{\max}$ denote, respectively, the minimal birth scale and maximal death scale within the corresponding ensemble.}. 
The dashed horizontal line indicates the termination scale of the persistence computation, at which only a single connected component remains, represented by the blue point on the line. The $H_0$ classes (blue points) disappear when connected components merge, giving rise to $H_1$ classes (orange points). These one-cycles subsequently die when they are filled in by higher-dimensional simplices, corresponding to the emergence of $H_2$ classes (green points).\footnote{In the full six-dimensional moduli space, we also observe short-lived $H_3$ and $H_4$ classes, which account for the disappearance of certain short-lived $H_2$ cycles through mergers into higher-dimensional features. A detailed resolution of this behaviour would require subsampling of the full dataset. By contrast, the computation of $H_1$ classes can be performed on the complete dataset~B, which is crucial for obtaining reliable results, albeit at the expense of a systematic exploration of higher-dimensional homology.}
Cycles with longer lifetimes correspond to larger-scale geometric structures in the complex planes; the most persistent $H_1$ class is marked by a cross highlighting the associated orange point. The vertically aligned orange points represent one-cycles that are born at the same filtration scale but persist for different durations. In particular, each individual moduli plane exhibits long-lived $H_1$ classes, signalling robust loop-like structures in the vacuum distribution. In the bottom-right panel of Fig.~\ref{fig:PH_roots_B}, no cycles appear at small scales ($\leq 0.18$), reflecting the effect of farthest-point sampling, which emphasises long-lived features and suppresses short-lived ones.

It is instructive to compare these findings with well-studied toy models, such as rigid Calabi-Yau compactifications and toroidal flux compactifications with $W_0=0$ and enhanced symmetries, analysed in \cite{Cirafici:2015pky,Cole:2018emh}. In those settings, the persistence diagrams for the rigid Calabi-Yau model display a larger number of long-lived $H_1$ classes. These persistent one-cycles correspond directly to extended empty regions in moduli space and admit a clear physical interpretation in terms of tadpole cancellation constraints and the rational structure of $\tau$, which is expressed as a ratio of flux quanta. 

Although a direct visualisation of the full six-dimensional moduli space is not possible, it is essential to perform the persistent homology analysis on the complete space. Such an analysis can reveal topological features that are invisible in lower-dimensional projections and therefore provides a more faithful description of the global structure of the vacuum distribution. In particular, certain long-lived $H_1$ classes observed in projected subspaces may disappear once the full six-dimensional geometry is taken into account. In the full moduli space, the resulting persistence diagrams are dominated by numerous short-lived $H_1$ and $H_2$ classes, represented by orange and green points in Fig.~\ref{fig:PH_roots_B}, respectively. These features are concentrated near the diagonal and are best interpreted as topological noise, reflecting the absence of pronounced local clustering rather than robust geometric structure. This behaviour stands in contrast to the projected analyses, where genuinely persistent one-cycles were identified.

\subsection{Persistent homology on flux space}
\label{sec:tda_results_flux}

We now analyse the persistent homology of the flux space associated with dataset~B. This study aims to uncover topological structures in flux space that arise from the combined constraints of tadpole cancellation and the \emph{ISD} condition. While the tadpole constraint alone does not restrict the fluxes to a compact region, it imposes a quadratic relation that, together with the linear ISD condition \eqref{eq:ISDCond_real}, significantly shapes the geometry on which the allowed flux configurations live.

\begin{figure}[h!]
    \centering
    \includegraphics[width=1.0\linewidth]{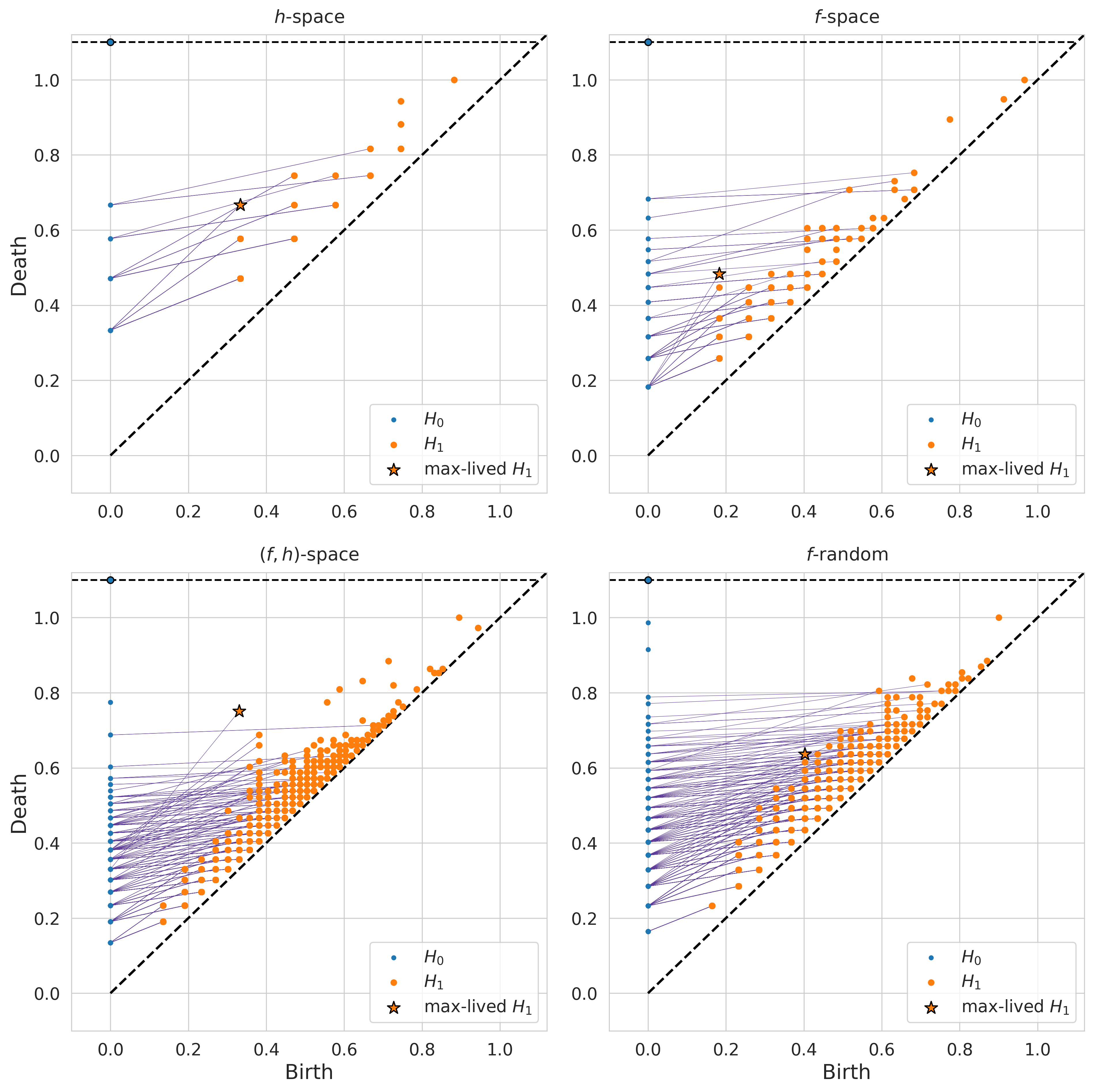}
    \caption{Persistent homology of the flux configurations in dataset~B: (top left) $h$-flux subspace, (top right) $f$-flux subspace, and (bottom left) the combined $(f,h)$ flux space. The $f$-flux subspace exhibits a larger number of $H_1$ classes (orange) and $H_0$ classes (blue), attributable to the broader range of flux quanta and the increased multiplicity of distinct flux configurations, leading to a more pronounced lattice-like organisation in flux space. Bottom right: Persistence diagram for a reference ensemble generated via empirical coordinate-wise sampling, preserving the marginal integer distributions of the $f$-flux coordinates while removing inter-coordinate correlations.}
    \label{fig:PH_fluxes_B}
\end{figure}

The corresponding persistence diagrams are shown in Fig.~\ref{fig:PH_fluxes_B} for four ensembles: $h$-fluxes (top left), $f$-fluxes (top right), combined $(f,h)$-fluxes (bottom left), and a randomised $f$-flux ensemble (bottom right). To facilitate a meaningful comparison across these ensembles, all finite birth--death pairs are rescaled to the interval $[0,1]$ as described previously. The maximal rescaled lifetimes of the $H_1$ classes are $0.333$, $0.300$, $0.420$, and $0.250$ for the $h$-flux, $f$-flux, $(f,h)$-flux, and random $f$-flux ensembles, respectively. To construct a statistically comparable random $f$-flux ensemble,\footnote{Sampling all six $f$-flux coordinates uniformly over a hyper-rectangular domain leads to a concentration of points near the boundary, which artificially increases the typical birth and death scales in the persistence diagrams. This renders a direct comparison with the physical $f$-flux ensemble less informative.} we proceed as follows. For each of the six $f$-flux coordinates, we determine the set of distinct integer values $k$ appearing in dataset~B and compute their empirical frequencies
\begin{equation}
   p_i(k) = \frac{n_i(k)}{\sum_{k'} n_i(k')}\,, 
\end{equation}
where $n_i(k)$ denotes the number of occurrences of the value $k$ in the $i$-th coordinate. These normalised frequencies define a discrete probability distribution for each coordinate. New random $f$-flux vectors are then generated by sampling each coordinate $i=1,\ldots,6$ independently according to its empirical distribution $p_i(k)$. This procedure preserves the marginal statistics of each flux coordinate while eliminating correlations between them, thereby providing a randomised reference ensemble that retains the dominant integer structure of the original $f$-flux data.

In Fig.~\ref{fig:PH_fluxes_B}, the persistence diagrams indicate a markedly larger number and greater lifetime of $H_1$ classes in the $f$-flux sector compared with the $h$-flux sector. This difference can be traced to the broader range and larger typical magnitudes of the $f$-flux quanta, which induce a wider spread in pairwise distances and therefore generate a more extended hierarchy in the filtration parameter. As a consequence, one-cycles in the $f$-flux ensemble persist over larger scale intervals. The combined $(f,h)$ ensemble exhibits a broad distribution of moderately long-lived $H_1$ classes and contains the most persistent feature among all four ensembles considered. By contrast, the $f$-random reference ensemble produces a dense concentration of short-lived cycles with comparatively small persistence. Overall, the $h$-flux and $f$-flux ensembles display more organised and hierarchical topological behaviour, whereas the $f$-random ensemble lacks this structured hierarchy.

A striking feature of the flux persistence diagrams is the appearance of vertically aligned sets of points at discrete birth scales, together with a repetition of the same pattern of vertical spacings across different birth values. Each vertical alignment corresponds to a collection of topological features born at an identical filtration scale, reflecting the discrete, lattice-like structure of the flux space. Although the separations between successive death times along a single vertical line are not uniform, the same sequence of separations recurs across multiple birth scales. This repetition indicates that the relative arrangement of features within a given birth scale is replicated at higher scales, consistent with the underlying integer lattice geometry. Furthermore, we observe a pronounced correlation between the death times of $H_0$ classes and the birth times of $H_1$ classes. In the diagrams in Fig.~\ref{fig:PH_fluxes_B}, these correlated events are indicated by purple line segments connecting a dying $H_0$ class to one or more $H_1$ classes born at the same filtration value. While persistence diagrams do not explicitly encode persistence pairings, these coincidences in filtration scale\footnote{The indicated connections between $H_0$ death times and $H_1$ birth times are inferred from matching filtration values in the persistence diagrams and have not been analysed at the level of explicit persistence pairings.} suggest a common geometric origin: at specific lattice spacings, the addition of simplices simultaneously merges previously disconnected components and closes loops. 

Taken together, these observations point to a layered and highly coordinated topological structure in flux space, induced by the integer quantisation of fluxes. The discrete lattice geometry governs both the hierarchical emergence of cycles and the correlated merging events observed in the persistence analysis. While persistent homology thus provides a global and scale-resolved characterisation of the topological organisation of flux vacua, it does not yield an explicit low-dimensional parametrisation of the data. To obtain such a representation, capable of compressing the high-dimensional flux configurations into a compact set of latent variables, we now turn to non-linear representation learning via autoencoders.

\section{Autoencoder latent representation}
\label{sec:autoencoders}

To further probe the structure of our datasets, we employ an autoencoder neural network that maps the 12-dimensional integer flux vectors to a two-dimensional latent representation. The purpose of this non-linear dimensionality reduction is not solely to achieve compression, but to learn a compact parametrisation that organises the flux configurations according to salient properties of the corresponding vacua. 

The autoencoder architecture consists of an encoder and a decoder network. In our implementation, the encoder takes an input flux vector $\mathbf{x}\in\mathbb{R}^{12}$ and maps it, via two hidden layers with non-linear $\mathrm{ReLU}$ activation functions, to a latent vector $\mathbf{z}\in\mathbb{R}^{2}$. This latent representation is then fed into a decoder network of comparable depth, which attempts to reconstruct the original flux vector, yielding an output $\hat{\mathbf{x}}\in\mathbb{R}^{12}$. The network is trained by minimising a reconstruction loss that penalises deviations between $\mathbf{x}$ and $\hat{\mathbf{x}}$.

To bias the latent space towards capturing features relevant for the vacuum structure, we augment the architecture with an additional feedforward network acting directly on $\mathbf{z}$. This auxiliary network is trained to predict the value of the superpotential, thereby encouraging the encoder to organise the latent variables in a manner correlated with $W_0$ (see \cite{Betzler:2020rfg,Krippendorf:2025mhp} for similar physical losses). In this way, the learned two-dimensional representation is shaped not only by geometric proximity in flux space, but also by the associated vacuum data.

The total training objective is designed to incorporate several physically motivated constraints and is defined as
\begin{equation}
    \mathcal{L}= \mathcal{L}_{\text{rec}} + \lambda_{1}\mathcal{L}_{W_0} + \lambda_{2}\mathcal{L}_{N_{\text{flux}}}  + \lambda_{3}\mathcal{L}_{\text{lat}}
\end{equation}
where the coefficients $\lambda_i$ are hyperparameters controlling the relative weight of each contribution.
The first term is the standard reconstruction error between the input and output fluxes, 
\begin{equation}
    \mathcal{L}_{\text{rec}}=\frac{1}{{\mathcal{N}_{\text{vac}}}}\sum^{\mathcal{N}_{\text{vac}}}_i|\hat{\text{\textbf{x}}}^i-\text{\textbf{x}}^i|^2\, .
\end{equation}
It penalises deviations between the input flux vectors $\mathbf{x}^{\,i}\in\mathbb{R}^{12}$ and their reconstructions $\hat{\mathbf{x}}^{\,i}$.
In addition, we include a term enforcing the correct value of the reconstructed superpotential $W_0(\hat{\text{\textbf{x}}}^i, \phi^i)$ obtained by the combination of the reconstructed fluxes with the original moduli $\phi \in \mathbb{R}$,
\begin{equation}
    \mathcal{L}_{W_0}=  \frac{1}{{\mathcal{N}_{\text{vac}}}}\sum_{i=0}^{\mathcal{N}_{\text{vac}}} |W_0(\hat{\text{\textbf{x}}}^i, \phi^i)-W_0(\text{\textbf{x}}^i, \phi^i)|^2 \, ,
    \label{eq:loss_W0}
\end{equation}
where $\phi^{\,i}\in\mathbb{R}^6$ denotes the moduli associated with the $i$-th vacuum.
We further impose approximate preservation of the tadpole constraint through
\begin{equation}
    \mathcal{L}_{N_{\text{flux}}}=  \frac{1}{{\mathcal{N}_{\text{vac}}}}\sum_{i=0}^{\mathcal{N}_{\text{vac}}}|N_{\text{flux}}(\hat{\text{\textbf{x}}}^i)-N_{\text{flux}}(\text{\textbf{x}}^i)|^2\, .
    \label{eq:loss_Nflux}
\end{equation}
Finally, we include a supervised loss acting directly on the latent representation,
\begin{equation}
    \mathcal{L}_{\text{lat}} = \frac{1}{{\mathcal{N}_{\text{vac}}}} \sum_{i=0}^{{\mathcal{N}_{\text{vac}}}}|\hat{W_0}(\text{z}^i)-W_0(\text{x}^i, \phi^i)|^2 \ ,
\end{equation}
where $\mathbf{z}^{\,i}\in\mathbb{R}^2$ is the encoded latent vector, $\hat{W}_0(\mathbf{z}^{\,i})$ is the superpotential predicted by the auxiliary latent-space network, and the $W_0(\text{x}^i, \phi^i)$ is the true value computed from the original fluxes and moduli. This term explicitly encourages the latent variables to encode information relevant for $W_0$. Notice that in \eqref{eq:loss_W0} and \eqref{eq:loss_Nflux} we round off the reconstructed fluxes $\hat{\textbf{x}}$. The combined loss is minimised using the Adam optimiser, with representative hyperparameter choices $\lambda_1=\lambda_3=10$ and $\lambda_2=0.1$, thereby prioritising accurate reconstruction of $W_0$ and its latent prediction while softly enforcing the tadpole constraint.

\begin{figure}[t!]
    \centering
    \includegraphics[width=1.0\textwidth]{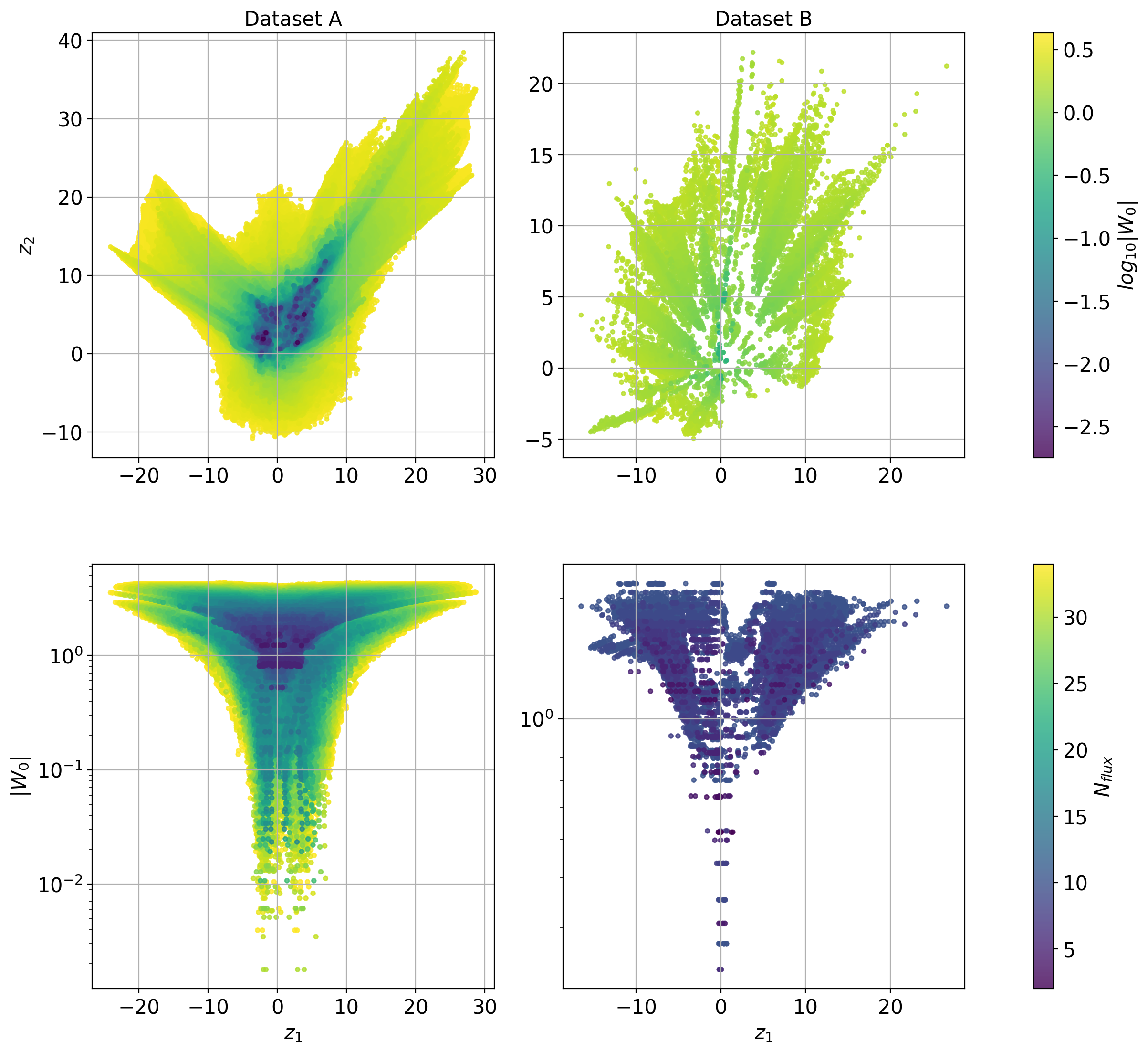}
    \caption{Visualisation of the learned latent representations and the associated superpotential structure for the two datasets. \textit{Top row}: two-dimensional latent space, with each point colour-coded by the value of the superpotential $W_0$. \textit{Bottom row}: one-dimensional projections displaying the relation between the latent coordinate $z_1$ and $W_0$, where points are coloured according to the corresponding flux-induced tadpole $N_{\mathrm{flux}}$.}
    \label{fig:Autoencoder}
\end{figure}

In Fig.~\ref{fig:Autoencoder}, we display the learned two-dimensional latent representation obtained after training. The vacua are not distributed uniformly in this plane; rather, they organise along well-defined directions and form distinct clusters. In particular, configurations with small values of the superpotential accumulate in a central region of the latent space. This behaviour indicates that the autoencoder has identified latent directions that correlate strongly with variations in $W_0$, thereby aligning relevant features of the flux landscape with the learned coordinates. This correlation becomes even more apparent in the one-dimensional projection onto the latent coordinate $z_1$. When plotting $W_0$ against $z_1$, vacua with progressively smaller values of the superpotential are seen to cluster around $z_1 \simeq 0$. Compared with the linear PCA analysis, this demonstrates a clearer organisation of the data, suggesting that the non-linear encoding captures structure that is not accessible through variance-based linear projections.

Note that clustering observed in a latent space does not, in general, imply clustering in the original space. Dimensionality reduction, whether linear (e.g. PCA) or nonlinear (e.g. autoencoders), can distort pairwise distances and geometric structure. While PCA provides a linear, variance-preserving projection with a transparent geometric interpretation, it does not preserve distances after truncation. Autoencoders, being non-linear and learned mappings, may introduce even stronger geometric distortions. Consequently, cluster structure in latent space should not automatically be interpreted as reflecting clustering in the original space.

Although the present study focuses on a specific ensemble of no-scale Type~IIB flux vacua, the methodology is not tied to this particular construction. The capacity of autoencoders to approximate low-dimensional manifolds and to isolate directions correlated with selected vacuum properties is readily transferable to more intricate and higher-dimensional compactifications, where the geometric and statistical structure of the solution space is expected to be substantially richer.

\section{Conclusion and Summary}
\label{sec:summary}

In this work, we investigated several strategies for parameter compression in string-theoretic datasets. Our analysis was based on the exhaustive dataset of Type~IIB flux vacua constructed in \cite{Chauhan:2025rdj}, which provides a complete enumeration of solutions within a fixed region of moduli space.
We performed linear and non-linear dimensionality reductions of the flux configurations to uncover features of the landscape highlighting the importance of physical losses on shaping the latent space embedding.
The principal component analysis demonstrated that the flux ensemble exhibits a pronounced concentration of variance: although embedded in a 12-dimensional space, the data are effectively organised along approximately five dominant directions. These principal directions correspond to specific linear combinations of flux quanta that account for the bulk of the statistical variation across the ensemble. Within the PCA-reduced space, vacua with small values of the flux superpotential $|W_0|$ are not homogeneously distributed. Instead, they cluster near the origin of the leading principal component and within a restricted region of the reduced coordinates. This localisation reflects the non-trivial constraints imposed by the combined ISD and tadpole conditions, which correlate different flux quanta in a structured manner.

Complementing the PCA analysis, we employed tools from TDA to extract robust topological features in our datasets, viewed as finite point clouds embedded in moduli and flux space. Using persistent homology, we identified long-lived cycles in the individual $\tau$, $z^1$, and $z^2$ subspaces, which are not apparent when considering the complete moduli space. This indicates that TDA can isolate persistent structures for the stabilised moduli whose scale is set by the tadpole bound on the corresponding flux configurations. In flux space, the persistence diagrams exhibit a highly regular pattern that reflects the underlying integer lattice structure imposed by flux quantisation. The discrete nature of the flux ensemble leads to repeated birth scales and correlated merger events in the filtration, providing a topological signature of the algebraic constraints defining the vacua. In this way, TDA complements the compression-based analysis by offering a global and multi-scale characterisation of the sampled flux landscape.

To capture non-linear correlations beyond the PCA results, we employed a physics-informed autoencoder to construct a low-dimensional latent representation of the flux space. By incorporating loss terms associated with the flux superpotential and tadpole conditions, the autoencoder learned a latent space representation that organises solutions according to given input criteria. In particular, we introduced a supervised term that encourages the latent variables to encode information predictive of $W_0$. As a consequence, the resulting latent space organises vacua according to their superpotential value: configurations with small $|W_0|$ are mapped to a sharply localised central region. This behaviour both sharpens and extends the correlation already visible in the leading PCA direction, demonstrating that the non-linear model captures structure that cannot be reduced to linear variance alone. More broadly, this framework is readily generalisable. The same architecture can be adapted to incorporate additional physical observables, alternative constraints, or larger flux ensembles, and can be applied to more intricate compactifications with higher-dimensional parameter spaces. In this sense, physics-informed representation learning provides a systematic avenue for extracting compressed yet physically structured parametrisations of the string landscape.

We emphasise that our analysis is based on exhaustive datasets of flux vacua and therefore does not rely on incomplete or randomly sampled subsets. In this sense, it provides a controlled setting for identifying latent representations that faithfully encode the structural and algebraic constraints defining the vacuum ensemble. This constitutes an important step towards constructing foundation models whose internal parametrisations reflect genuine features of the flux landscape rather than artefacts of limited sampling.

The methodology we have developed is sufficiently general to be extended to larger regions of moduli space, such as the expanded datasets presented in \cite{Chauhan:2025rdj}, provided suitable subsampling or approximation strategies are implemented to manage computational complexity. Such extensions would enable systematic cross-dataset comparisons, allowing one to assess the robustness and potential universality of the correlations and topological features identified here. Preliminary comparisons among the ensembles studied in this work already suggest that several of the observed correlations and persistent structures remain stable under enlargement of the moduli domain. This stability indicates that the extracted patterns are unlikely to be artefacts of a particular sampling window and instead point towards more intrinsic organisational principles of the flux landscape.

\section*{Acknowledgments}

We thank Arthur Hebecker, Magdalena Larfors, Zhimei Liu, Gary Shiu, Gaetano Sifo, and Moritz Walden for interesting discussions. SK has been partially supported by STFC consolidated grants ST/T000694/1 and ST/X000664/1. AM would like to thank the Leinweber Institute for Theoretical Physics for supporting his sabbatical visit at the University of Michigan, Ann Arbor. The work of PP is supported by a NYUAD research grant. The research of AS is supported by NSF grant PHY-2309456.
We want to thank the creators of the Python library \texttt{Ripser} \cite{ctralie2018ripser}. This work made use of the open-source software 
\texttt{CYTools}~\cite{Demirtas:2022hqf},
\texttt{jax}~\cite{jax2018github},  
\texttt{matplotlib}~\cite{Hunter:2007},  
\texttt{numpy}~\cite{harris2020array1},
\texttt{scikit-learn}~\cite{scikit-learn},
\texttt{scipy}~\cite{2020SciPy-NMeth1},
and \texttt{seaborn}~\cite{Waskom2021}. 


\appendix

\section{PCA Details }
\label{app:pca_comp}

\subsection*{Principal components of the flux space}

The algorithmic procedure underlying PCA is outlined in Algorithm~\ref{algorithm:PCA}. We emphasise that the optional rescaling of each feature by its standard deviation, implemented to normalise the dataset, should be guided by the specific aims of the analysis, as it directly affects the relative weight assigned to variance in different directions.  

\begin{algorithm}[htb!]
\caption{Principal Component Analysis}
\begin{algorithmic}[A]
\Require Data matrix $X \in \mathbb{R}^{n \times d}$ with $n$ samples and $d$ features
\Ensure Reduced data matrix $Z \in \mathbb{R}^{n \times k}$, with $k \leq d$ principal components
\State \textbf{Standardise the data:} For each feature $j=1,\dots,d$ compute
    \[
    \mu_j = \frac{1}{n} \sum_{i=1}^{n} X_{ij}, 
    \quad
    \sigma_j = \sqrt{\frac{1}{n} \sum_{i=1}^{n} (X_{ij} - \mu_j)^2}\, ,
    \]
    and form the standardised matrix
    \[
    \tilde{X}_{ij} = \frac{X_{ij} - \mu_j}{\sigma_j}\, .
    \]
\State \textbf{Compute the covariance matrix:}
    \[
    \Sigma = \frac{1}{n} \tilde{X}^\top \tilde{X} \in \mathbb{R}^{d \times d}
    \]
\State \textbf{Compute the eigenvalues and eigenvectors of } $\Sigma$
    \[
    \Sigma v_i = \lambda_i v_i\, .
    \]
\State \textbf{Sort eigenvectors} $v_i$ by decreasing eigenvalues $\lambda_i$
\State \textbf{Select the top $k$ eigenvectors} to form projection matrix $W \in \mathbb{R}^{d \times k}$
\State \textbf{Project the data:}
    \[
    Z = \tilde{X} W\, .
    \]
\end{algorithmic}
\end{algorithm}\label{algorithm:PCA}

The first three principal components of the flux space for datasets~A and~B are presented in Tables~\ref{tab:pc_flux_A} and~\ref{tab:pc_flux_B}. The explicit expressions demonstrate that the individual flux quanta do not contribute uniformly to the dominant variance directions. For instance, the leading component (PC1) is primarily driven by $(f_1)^2$, while PC2 receives its largest contribution from $(h_1)^2$. We further observe that $(h_2)^1$ appears only in the final principal component. This is consistent with the fact that $(h_2)^1=0$ for all vacua in the datasets, and hence does not contribute to the variance of the flux distribution.

\begin{table}[h!]
\centering
\renewcommand{\arraystretch}{1.1}
\resizebox{\textwidth}{!}{
\begin{tabular}{r|cccccccccccc}
PC & $(f_1)^1$ & $(f_1)^2$ & $(f_1)^3$ & $(f_2)^1$ & $(f_2)^2$ & $(f_2)^3$ & $(h_1)^1$ & $(h_1)^2$ & $(h_1)^3$ & $(h_2)^1$ & $(h_2)^2$ & $(h_2)^3$ \\ \hline
PC1& 0.069 & 0.920 & 0.371 & 0.001 & 0.016 & 0.002 & 0.086 & 0.020 & -0.009 & 0.000 & 0.007 & 0.056 \\ 
PC2& 0.280 & -0.015 & 0.044 & 0.001 & 0.123 & -0.148 & -0.059 & -0.926 & -0.146 & 0.000 & -0.005 & -0.023 \\
PC3& 0.833 & 0.038 & -0.247 & 0.001 & -0.082 & 0.444 & -0.039 & 0.174 & -0.087 & 0.000 & 0.002 & -0.029 \\

\end{tabular}}
\caption{Principal components of the flux space for dataset~A.}
\label{tab:pc_flux_A}
\end{table}

\begin{table}[h!]
\centering
\renewcommand{\arraystretch}{1.1}
\resizebox{\textwidth}{!}{
\begin{tabular}{r|cccccccccccc}
PC & $(f_1)^1$ & $(f_1)^2$ & $(f_1)^3$ & $(f_2)^1$ & $(f_2)^2$ & $(f_2)^3$ & $(h_1)^1$ & $(h_1)^2$ & $(h_1)^3$ & $(h_2)^1$ & $(h_2)^2$ & $(h_2)^3$ \\ \hline
\text{PC1} & 0.0479 & 0.9194 & 0.3868 & 0.0126 & 0.0095 & 0.0082 & -0.0452 & 0.0132 & -0.0049 & 0.0000 & 0.0024 & 0.0201 \\
\text{PC2} & 0.8629 & 0.0549 & -0.2295 & -0.0090 & -0.0478 & 0.4045 & 0.0678 & -0.1487 & -0.0824 & 0.0000 & 0.0011 & -0.0134 \\
\text{PC3} & -0.0515 & 0.0420 & 0.0234 & -0.0777 & 0.0059 & -0.0285 & 0.9927 & 0.0487 & 0.0162 & 0.0000 & 0.0066 & 0.0058 \\

\end{tabular}}
\caption{Principal components of the flux space for dataset~B.}
\label{tab:pc_flux_B}
\end{table}

\section{TDA on dataset~A}
\label{app:tda_datA}

\begin{figure}[ht]
    \centering
    \includegraphics[width=1.0\linewidth]{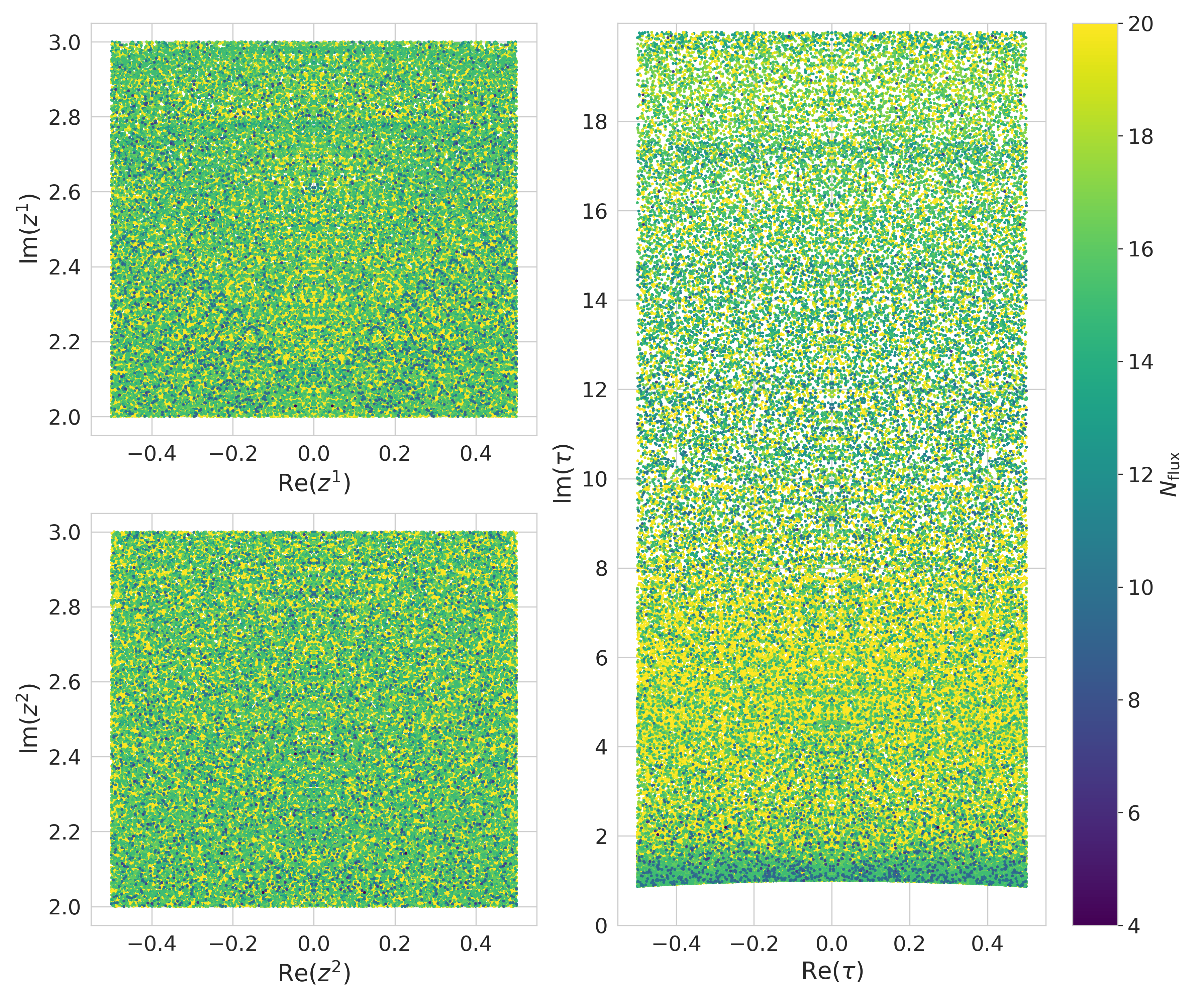}
    \caption{Distributions of the moduli VEVs with flux-induced D3-brane charge $N_{\text{flux}}\leq 20$ in dataset~A (recall Tab.~\ref{tab:summary}): (top left) projection onto the $z^1$-plane, (bottom left) projection onto the $z^2$-plane, and (right) projection onto the $\tau$-plane. In each case, the distributions are symmetric under reflection about the imaginary axis and exhibit non-trivial geometric structures, including arc-like features and closed loops. The complex structure moduli $z^1$ and $z^2$ display qualitatively distinct patterns, reflecting the asymmetric role they play in the geometry of the underlying Calabi-Yau manifold $\mathbb{CP}^4_{[1,1,1,6,9]}$ along the corresponding moduli directions.
    }
    \label{fig:roots_A}
\end{figure}

For dataset~A, the VEVs of the complex structure moduli $z^i$ and the axio-dilaton $\tau$ are shown in Fig.~\ref{fig:roots_A}. In this case, constructing Vietoris--Rips simplicial complexes directly on the full set of vacua is computationally prohibitive due to the large number of points.\footnote{An alternative strategy is to use \emph{witness complexes} \cite{10.5555/2386332.2386359}, which reduce computational cost by selecting a sparse set of landmark points while approximately preserving the topological structure of the full dataset. Concretely, one selects a subset of landmarks $L \subset X$ from the complete point cloud $X$. Simplices are formed only among the landmarks, while the remaining points act as \emph{witnesses}. A $k$-simplex spanned by $k+1$ landmarks is included if there exists at least one witness whose distances to these landmarks are collectively small relative to its distances to other landmarks. This construction significantly reduces the number of simplices and provides a controlled approximation to the persistent homology of the Vietoris--Rips or \v{C}ech complexes built on $X$.}

\begin{algorithm}[ht]
\caption{Farthest Point Sampling (FPS)}
\begin{algorithmic}[1]
\Require Point set $X = \{x_1, \dots, x_N\} \subset \mathbb{R}^d$, number of samples $k$
\Ensure Sampled set $S \subset X$, $|S| = k$
\State Initialise $S \gets \emptyset$, $d_i \gets \infty$ for all $i$
\State Choose a random initial point $p_0 \in X$, add to $S$
\For{$i = 1$ to $N$}
    \State $d_i \gets \|x_i - p_0\|$
\EndFor
\For{$j = 2$ to $k$}
    \State $x_{\text{farthest}} \gets \arg\max_i d_i$
    \State $S \gets S \cup \{x_{\text{farthest}}\}$
    \For{$i = 1$ to $N$}
        \State $d_i \gets \min(d_i, \|x_i - x_{\text{farthest}}\|)$
    \EndFor
\EndFor
\State \Return $S$
\end{algorithmic}
\end{algorithm}\label{algorithm:FPS}

 To make the computation tractable, we restrict the analysis to the subset of vacua with $N_{\text{flux}}\leq 20$ and perform farthest point sampling before constructing the Vietoris--Rips complex. This greedy algorithm constructs a representative subset of a metric point cloud by iteratively selecting the point farthest from the current sample. At each step, the chosen point maximises its minimal distance to the previously selected points, thereby ensuring broad geometric coverage of the dataset. The corresponding algorithm is provided in Algorithm~\ref{algorithm:FPS} .

\begin{figure}[ht]
    \centering
    \includegraphics[width=0.9\linewidth]{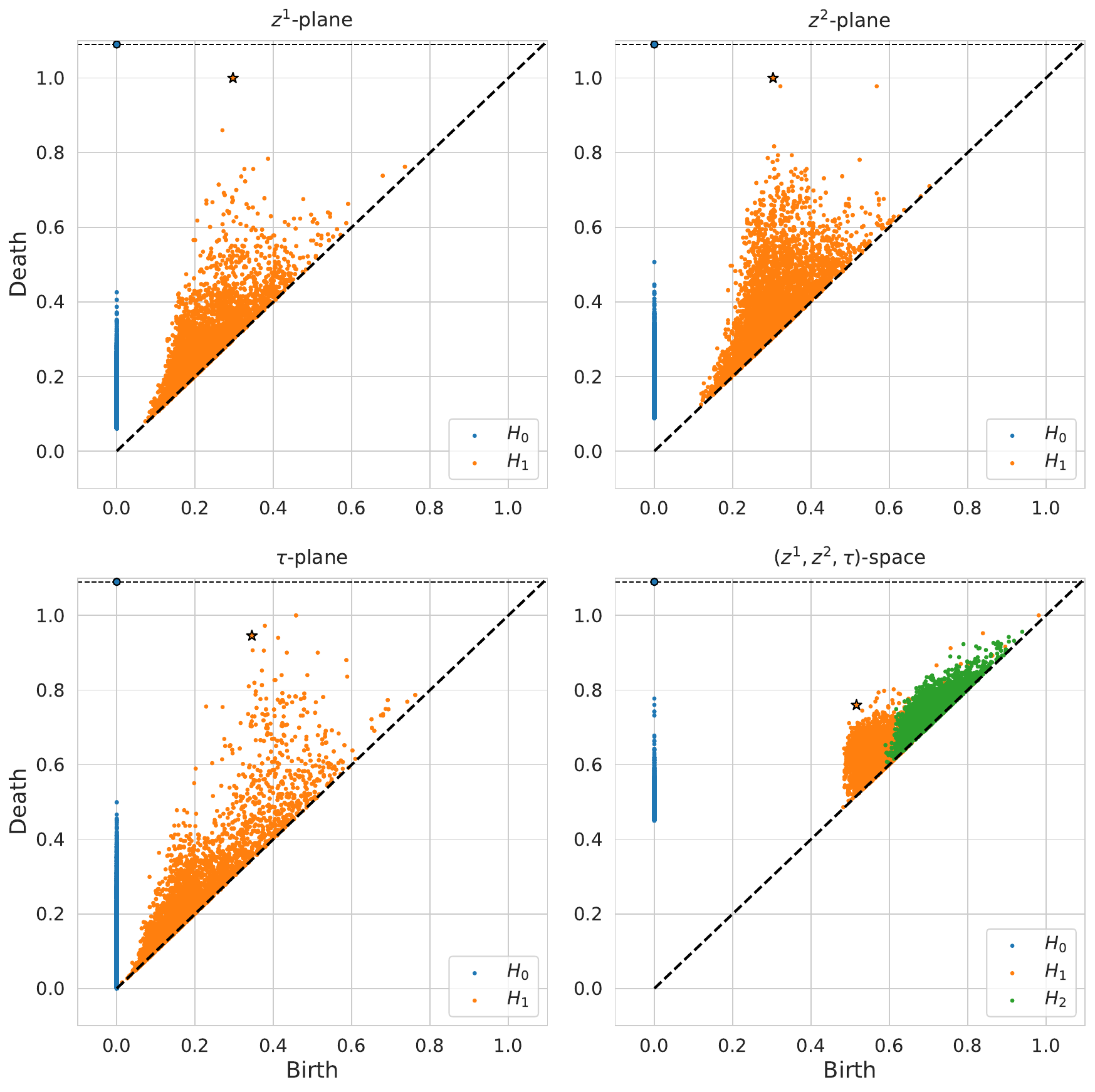}
\caption{Persistent homology analysis of vacua in dataset~A ( $N_{\text{max}}\leq 20$, with farthest point sampling): (top left) $z^1$-plane, (top right) $z^2$-plane, (bottom left) $\tau$-plane, and (bottom right) the complete 6-dim $(z^1, z^2, \tau)$-space. 0-cycles are shown in blue and 1-cycles in orange, with the most persistent 1-cycle marked by a cross. Several long-lived 1-cycles are visible in the projected planes, whereas no comparably persistent features appear in the full moduli space. Additionally, the existence of 2-cycles shown in green is also analysed for the complete moduli space.}
    \label{fig:PH_roots_A}
\end{figure}

\begin{figure}[ht]
    \centering
    \includegraphics[width=0.9\linewidth]{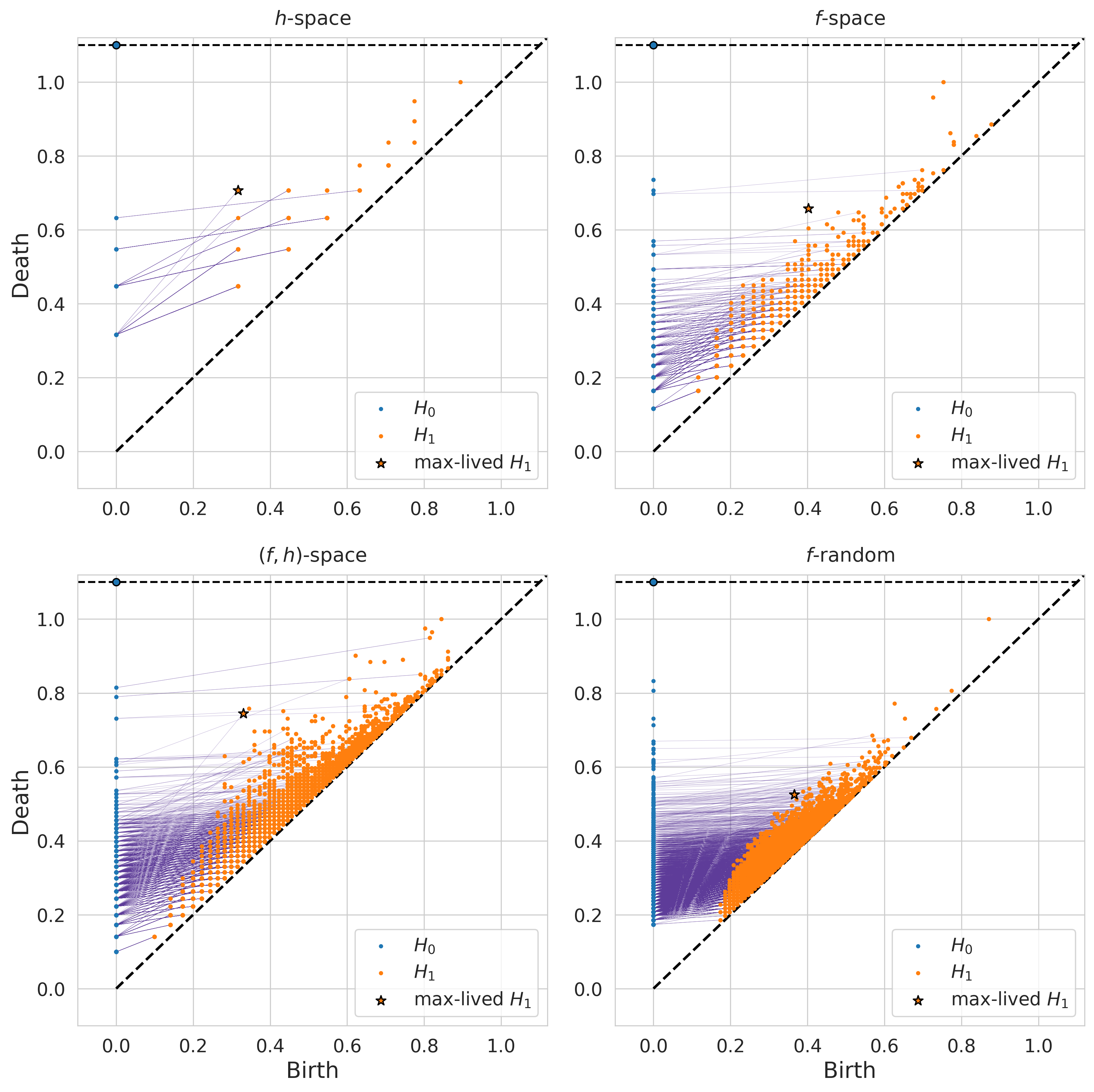}
\caption{Persistent homology analysis of fluxes in dataset~A ($N_{\text{flux}}\leq 20$, with farthest point sampling): (top left) $h$-flux subspace, (top right) $f$-flux subspace, and (bottom left) the combined $(f,h)$ flux space. The $f$-flux subspace exhibits a larger number of $H_1$ classes (orange) and $H_0$ classes (blue), attributable to the broader range of flux quanta and the increased multiplicity of distinct flux configurations, leading to a more pronounced lattice-like organisation in flux space. Bottom right: Persistence diagram for a reference ensemble generated via empirical coordinate-wise sampling, preserving the marginal integer distributions of the $f$-flux coordinates while removing inter-coordinate correlations.}
    \label{fig:PH_fluxes_A}
\end{figure}

This procedure successfully reproduces the dominant long-lived cycles present in the full dataset\footnote{Here, “full dataset” does not refer to the entirety of dataset~A (for which $N_{\text{flux}}\leq 34$), but rather to the subset on which the farthest-point sampling is performed, namely configurations satisfying $N_{\text{flux}}\leq 20$.}. The corresponding persistence diagrams for the moduli VEVs and flux configurations of dataset~A are displayed in Fig.~\ref{fig:PH_roots_A} and Fig.~\ref{fig:PH_fluxes_A}, respectively. Note that to have meaningful comparison with the results for dataset~B, we once again scale the lifetimes of various cycles in the interval $[0,1]$. The most persistent one-cycle is indicated by a $\star$.

For the moduli VEVs of dataset~A, we again observe numerous long-lived $H_1$ classes, in close analogy with dataset~B. The accumulation of long-lived cycles within similar ranges of birth and death scales signals the existence of several cycles of comparable size, reflecting repeated structural motifs in the distribution of vacua. Notice again that in the bottom-right panel of Fig.~\ref{fig:PH_roots_A} there are no cycles at small scales ($\leq 0.45$). As in dataset~B, this is a consequence of farthest-point sampling, which preferentially captures dominant long-lived topological features while suppressing short-lived (noisy) ones.

For the flux configurations, the persistence diagrams exhibit the same qualitative features identified for dataset~B, most notably the lattice-like organisation induced by flux quantisation. The contrast with the persistence diagram obtained from randomly generated $f$-flux configurations is even more pronounced in this case. As before, the vertically aligned orange points correspond to $H_1$ classes born at identical filtration values but possessing different lifetimes, indicating families of cycles emerging at common lattice scales. A notable difference relative to dataset~B, present in both the moduli and flux spaces, is the increased presence of short-lived cycles. These features arise from the larger number of vacua at high values of $N_{\mathrm{flux}}$, which generate finer-scale geometric structure in the point cloud and consequently produce additional short persistence intervals in the diagrams.

\newpage
\mbox{}
\newpage
\mbox{}
\newpage

\bibliographystyle{utphys}
\bibliography{biblio}

\end{document}